\documentclass[letterpaper, 10 pt, conference]{ieeeconf}  

\let\proof\relax 
\let\endproof\relax 
\usepackage{booktabs}
\pdfoutput=1
\usepackage{bbm}
\usepackage{mathptmx}
\usepackage{graphicx} 
\usepackage{amssymb}
\usepackage{amsthm,amsfonts,bm}
\usepackage{amsmath,mathtools}
\usepackage{graphicx} 
\usepackage{caption,cite}
\usepackage{tikz}
\usetikzlibrary{arrows}

\newtheorem{definition}{Definition}

\newtheorem{lemma}{Lemma}

\newtheorem{assumption}{Assumption}

\newtheorem{proposition}{Proposition}

\newtheorem{theorem}{Theorem}

\IEEEoverridecommandlockouts                              

\title{\LARGE \bf 
Implementability of Honest Multi-Agent Sequential Decision-Making with Dynamic Population
}

\author{Tao Zhang and Quanyan Zhu$^\dagger$
\thanks{$^\dagger$ Both authors are with the Department of Electrical and Computer Engineering, New York University, Brooklyn, New York, USA, 11201, Email: {\tt\small \{tz636, qz494\}@nyu.edu}}
}

\begin{document}

\maketitle
\thispagestyle{empty}
\pagestyle{empty}

\begin{abstract}

We study the design of decision making mechanism for resource allocations over a multi-agent system in a dynamic environment. 
%
Agents' privately observed preference over resources evolves over time and the population is dynamic due to adoption of stopping rules.
The proposed model designs the rules of encounter for agents participating in the dynamic mechanism by specifying an allocation rule and three payment rules to elicit agents' coupled decision makings of honest preference reporting and optimal stopping over multiple periods. 
The mechanism provides a special posted-price payment rule that depends only on each agent's realized stopping time to directly influence the population dynamics.
%
This letter focuses on the theoretical implementability of the rules in perfect Bayesian Nash equilibrium and characterizes necessary and sufficient conditions to guarantee agents' honest equilibrium behaviors over periods.
We provide the design principles to construct the payments in terms of the allocation rules and identify the restrictions of the designer's ability to influence the population dynamics.
The established conditions make the designer's problem of finding multiple rules to determining an optimal allocation rule.

\end{abstract}

\section{Introduction}

Multi-agent sequential decision making is an important research agenda in engineering (e.g., \cite{moradi2016state}), economics (e.g., \cite{kaihara2003multi}), and Artificial Intelligence (AI) (e.g., \cite{tuyls2012multiagent}).
In many scenarios, agents act on behalf of individuals or organizations and the multi-agent system needs to represent their preferences, embody the knowledge about the environment, and behave according to their rationality in decision makings.
%
Mechanism design approaches have been used to design the rules of encounters for the multi-agent systems to elicit actions, allocate resources, manage the cooperation, and rule the competition in heterogeneous environment where each agent is self-interested (e.g., \cite{ephrati1991clarke,vytelingum2010agent}).

This letter considers a dynamic multi-agent resource allocation mechanism, in which there is one central planner (CP) that aims to allocate resources periodically to agents based on their private preferences.
The mechanism is characterized by an allocation rule and three payment rules that specify allocation and payment to each agent at each period based on their preference.
The multiple self-interested agents strategically report their preferences to the CP and choose a time to stop over periods.
Each agent's preference evolves over time due to the learning-by-doing, which captures the effects of experiences (i.e., historical preferences and allocations) on agent's preference. 
The population is dynamic due to agent's stochastic departure by the adoption of the stopping rule.
To directly influence the population dynamics, one of the payment rules is a posted-price rule that depends only on the realized stopping time of each agent.
%

We are interested in a class of direct mechanisms in which agents are incentivized to truthfully report their true preferences (i.e., incentive compatibility) and focus on the theoretical implementability of the mechanism when each agent strategically makes the reporting and the stopping decisions simultaneously.
Revelation principle allows the CP to focus on the direct mechanisms to replicate equilibrium outcomes of indirect mechanisms (see, e.g., \cite{VickreyCounterspeculation}).
Agents' adoption of stopping rule, however, complicates the guarantee of implementability in the dynamic environment due to the coupling of agents' reporting strategy and stopping decision in the strategic reasoning about their self interests.
The consideration of behavior elicitations when each agent makes simultaneous multiple decisions that depend on their private information 
over periods distinguishes this work from other mechanism design problems that consider only the reporting decisions (e.g., \cite{pavan2009dynamic,athey2013efficient}) and optimal stopping problems that involve no truthful revealing of private information (e.g., \cite{lovasz1995efficient, lundgren2010optimal}).

We introduce the notion of honesty elicitation (HE) constraint that guarantees the incentive compatibility and elicits optimal stopping behaviors of agents.
The honesty of the reporting can obviate agents' strategic reasoning and simplify the dynamic decision makings.
At each period, the HE constraint requires a guarantee of no profitable deviations from honesty at current period as well as agents' planned future behaviors.
This multi-dimensional analysis is simplified by establishing a one-shot deviation principle that provides necessity and sufficiency of single-period truthfulness to imply multi-period honesty. %
We provide necessary and sufficient conditions to guarantee the implementability of the HE constraint in perfect Bayesian Nash equilibrium (PBNE-HI) and propose design principles to construct the payment rules in terms of the allocation rule.
Despite its independence of agent' preference, we observe that the ability of the CP to control the population dynamics is limited due to the design restrictions of the posted-price payment rule in PBNE-HI mechanism.

We are interested in multi-agent sequential decision-making problems, in which self-interested agents want to obtain a resource that is allocated by a third party, called central planner (CP, she).
We consider a dynamic population environment, in which the population evolves over time due to each agent's local decision of whether to continue or not by adopting an optimal stopping rule.
The resource is allocated among the agents periodically.
Time is discrete and indexed by $t\in \mathbb{T}=\{0,\dots,T\}$.
Suppose there are $n$ agents entering the resource allocation model. Let $I=\{1,\dots, n\}$ be the set of agent indices. The CP is indexed as agent $0$.

\textbf{Information structure.} The information structure of the environment is described by a tuple $<\mathbf{X}, \mathbf{F}_{0}>$.
Here, $\mathbf{X} = \times_{i\in I, t\in \mathbb{T}}X_{i,t}$ is the joint set of agent's preferences, with $X_{i,t}\equiv[\underline{x}_{i,t}, \bar{x}_{i,t}]$, with $-\infty<\underline{x}_{i,t}<\bar{x}_{i,t}<\infty$, as the period-$t$ set of agent $i$'s preference, and $\mathbf{F}_{0}=\{F_{i,0}\}_{i\in I}$ is the set of distributions of agent's initial preferences.
We assume that each agent's preference is independent from each other at each period.
Each agent $i$'s period-$t$ preference $x_{i,t}$ is his own private information that is unobservable to all other agents and the CP.
Each agent $i$ (he) report his period-$t$ preference by using a reporting strategy, $\alpha_{i,t}:X_{i,t}\mapsto X_{i,t}$, such that $\Tilde{x}_{i,t}=\alpha_{i,t}(x_{i,t})$ is reported to the CP given $x_{i,t}$ as his true preference.
Meanwhile, agent $i$ decides whether to stop or continue by adopting a stopping time rule $\zeta_{i}$, which will be elaborated later in the letter.
Notationally, let $x_{i}=\{x_{i,t}\}_{t\in \mathbb{T}}$, $x^{t}_{i} = \{x_{i,s}\}^{t}_{s=0}$, and $\tilde{x}$ is treated analogously. 
%
%

\textbf{Take-it-or-leave-it.} At the ex-ante stage, i.e., prior to $t=0$, the CP provides a take-it-or-leave-it offer to agents by specifying $<\sigma_{i,t}, \rho_{i,t}, \gamma_{i,t}, \beta_{i,t}>$ to each agent $i$, for each $t\in\mathbb{T}$.
Let $A_{i,t}$ be the set of period-$t$ resource for agent $i$.
Here, $\sigma_{i,t}: X_{i,t} \times X_{-i,t} \mapsto A_{i,t}$ is the period-$t$ \textit{allocation rule} for agent $i$, such that he receives $a_{i,t}=\sigma_{i,t}(\tilde{x}_{i,t}, \tilde{x}_{-i,t})$ when he reports $\tilde{x}_{i,t}$ and others $\tilde{x}_{-i,t}$.
$\rho_{i,t}: X_{i,t}\mapsto \mathbb{R}$ is agent $i$'s period-$t$ \textit{payment rule}, such that $p_{i,t}= \rho_{i,t}(\tilde{x}_{i,t}, \tilde{x}_{-i,t})$ is paid to him when he decides to continue, given his report $\tilde{x}_{i,t}$ and others' $\tilde{x}_{-i,t}$.
$\gamma_{i,t}:X_{i,t}\times X_{-i,t}\mapsto \mathbb{R}$ is the \textit{terminal payment rule} and $\beta_{i,t}: \mathbb{T}\mapsto \mathbb{R}$ is the \textit{posted-price terminal payment rule} that specify the payment $p_{i,t} = \gamma_{i,t}(\tilde{x}_{i,t}, \tilde{x}_{-i,t}) + \beta_{i,t}(t)$ for agent $i$ at $t$ when he reports $\tilde{x}_{i,t}$, decides to stop, and others report $\tilde{x}_{-i,t}$.
The posted-price $\beta_{i,t}$ is independent of agent's preference.

\textbf{Learning-by-doing.} 
Each agent updates his preference at the beginning of each period due to his learning-by-doing 
that leads to stochastic dynamics of the preference. 
The environment is dynamic due to agents' learning-by-doing and the population evolutions.
%
%
%
We assume that the dynamics of agent $i$'s preference are governed by a Markovian stochastic process uniquely characterized by $\sigma_{i}$, $\alpha_{i}$, and a set of \textit{transition kernels} $K_{i}=\{K_{i,t}\}_{t\in\mathbb{T}}$, where $K_{i,t}: X_{i,t-1}\times A_{i,t-1}\mapsto \Delta(X_{i,t})$ (here, $\Delta(X)$ denotes the set of probability measures over $X$.).
Denote agent $i$'s \textit{stochastic process} by $\Gamma_{i}[\sigma_{i}, \alpha_{i}, K_{i}]$.
Let $F_{i,t}(\cdot| x_{i,t-1}, a_{i,t-1}): X_{i,t}\mapsto [0,1]$ be the cumulative distribution function (CDF) corresponding to $K_{i,t}$, with density function $f_{i,t}$.
We assume $f_{i,t}(x_{i,t})>0$, for all $x_{i,t}\in X_{i,t}$, $i\in I$, $t\in\mathbb{T}$.
Given any $x^{t}_{i}$, 
we define agent $i$'s \textit{interim process} $\Gamma_{i,t}[\sigma_{i},\alpha_{i}, K_{i}; x^{t}_{i}]$ that consists of $t$ periods of realized $x^{t}_{i}$ and stochastic process starting from $t+1$ uniquely defined by $\sigma_{i},\alpha_{i}, K_{i}$, and $x^{t}_{i}$.
%
%
%
%
Let $\omega_{i,t}$ be uniformly distributed from $(0,1)$. 
Let $\kappa_{i,t}$ be defined as follows:
\begin{equation}\label{eq:dynamic_representation_0}
\setlength{\abovedisplayskip}{1pt}
\setlength{\belowdisplayskip}{1pt}
    \kappa_{i,t}(x_{i,t-1},a_{i,t-1}, \omega_{i,t}) = F^{-1}_{i,t}(\omega_{i,t}| x_{i,t-1}, a_{i,t-1}),
\end{equation}
where $F^{-1}_{i,t}$ is the inverse of $F_{i,t}$ defined as $F^{-1}_{i,t}(\omega_{i,t}| x_{i,t-1}, a_{i,t-1}) = \inf\{x_{i,t}: F_{i,t}(x_{i,t}|x_{i,t-1},a_{i,t-1})\geq \omega_{i,t}\}$.
Let $\hat{x}_{i,t}$ denote the random variable and $x_{i,t}$ is one sample of $\hat{x}_{i,t}$ and define $\hat{\omega}_{i,t}$ analogously.
Hence, the dynamics of $x_{i,t}$ can be described by $\kappa_{i,t}$ as, given $x_{i,t-1}$ and $a_{i,t-1}$,
\begin{equation}\label{eq:dynamic_representation_1}
\setlength{\abovedisplayskip}{1pt}
\setlength{\belowdisplayskip}{1pt}
    \hat{x}_{i,t} = \kappa_{i,t}(x_{i,t-1}, a_{i,t-1}, \hat{\omega}_{i,t}),
\end{equation}
i.e., if $\omega_{i,t}$ is uniformly distributed over $(0,1)$, then $\hat{x}_{i,t}$ is distributed according to $K_{i,t}(x_{i,t-1}, a_{i,t-1})$.
The population evolution is another factor that causes the dynamics, due to agents' adoption of a stopping rule that leads to a weakly decreasing population.
Each agent $i$ decides whether to stop or not at each period after observing his preference and sends his stopping decision with his report to the CP.
%
%
%

\textbf{Conventions.}
To describe the dynamic population, let $I_{t}=\{1,\dots, n_{t}\}$, with $1\leq n_{t}\leq n_{0}=n$ be the set of period-$t$ participating agent indices.
%
%
Let $e_{t}$ represent general agent index in $I_{t}$ and let $i_{t}\in I_{t}$ denote agent $i$'s ($i\in I_{0}=I$) period-$t$ index.
Let $\bm{\sigma}=\{\sigma_{i,t}\}_{i\in I, t\in\mathbb{T}}$, $\bm{\sigma}_{t} = \{\sigma_{i,t}\}_{i\in I}$, and $\sigma_{i} = \{\sigma_{i,t}\}_{t\in \mathbb{T}}$; 
Analogous treatments can be conducted for $A$, $\alpha$, $\rho$, $\gamma$, and $\beta$.
To simplify the notation, we omit the time index $t$ in subscript, e.g., $x_{i_{t}} \equiv x_{i_{t},t}$.
Let $\mathbb{T}_{t,t+k} = \{t,t+1,\dots, t+k\}$ with $\mathbb{T}_{t}=\mathbb{T}_{t, T}$.
We only show the allocation rule $\sigma$ and reporting strategy $\alpha$, i.e., $\mathbb{E}_{t}\big[\cdot|\alpha_{i}, \sigma_{i} \big]$ (resp. $\mathbb{E}\big[\cdot|\alpha_{i}, \sigma_{i} \big]$), to represent the expectation taken under the process $\Gamma_{i,t}[\sigma_{i},\alpha_{i}, K_{i};x^{t}_{i}]$ (resp. $\Gamma_{i}[\sigma_{i},\alpha_{i}, K_{i}]$). 
Table \ref{table:notations} lists the notations.
%


\begin{table}[htbp]
	\centering
	\caption{Summary of Notations}
	\begin{tabular}{c|c}
		\toprule  
		$x_{i_{t}}\in X_{i_{t}}$ & agent $i_{t}$'s period-$t$ preference \\ 
		\midrule 
		$\alpha_{i_{t}}$ & agent $i_{t}$'s reporting strategy \\
		\midrule 
		 $\sigma_{i_{t}}$& allocation rule for agent $i_{t}$ at $t$\\
		 \midrule 
		 $\rho_{i_{t}}$ & intermediate payment rule\\
		 \midrule 
		 $\gamma_{i_{t}}$, $\beta_{i_{t}}$ & terminating payment rules\\
		 \midrule 
		 $K_{i_{t}}$ & transition kernel\\
		 \midrule 
		 $\mathbb{E}_{t}[ \cdot | \alpha_{i}, \sigma_{i} ]$& period-$t$ interim expectation given $\alpha_{i}$, $\sigma_{i}$ \\
		 \midrule 
		 $\mathbb{E}[\cdot|\alpha_{i}, \sigma_{i}]$ & ex-ante expectation given  $\alpha_{i}$, $\sigma_{i}$\\
		 \midrule 
		 $C_{i}$ and $C_{i_{t}}$ & agent $i$'s ex-ante and period-$t$ interim expected payoff\\
		 \midrule 
		 $\zeta^{*}_{i_{t}}$ & optimal stopping rule\\
		 \midrule 
		  $\bm{\tau}^{*} = \{\tau^{*}_{i_{t}}\}$ & set of first-passage time\\
		  \midrule 
		  superscript $S$ or $\bar{S}$ & stop or continue\\
		  \midrule 
		  $d^{Y}_{i_{t}}(\cdot, \cdot; \sigma_{i})$ &  distance function, $Y\in \{S, \bar{S}\}$ \\
		  \midrule 
		  $\phi^{Y}_{i_{t}}(\cdot|\sigma_{i})$ & characterizing function, driving the PBNE-HI analysis\\
		\bottomrule  
	\end{tabular}
\end{table}\label{table:notations}


\section{Decision-Making Model}


%

Let $\bm{\tau}_{t}=\{\tau_{e_{t}}\}_{e_{t}\in I_{t}}$ be any joint set of time horizon of agents participating at period $t$.
%
%
Let $u_{i_{t}}: X_{i_{t}} \times A_{i_{t}}$ be the instantaneous utility (utility) function such that $u_{i_{t}}(x_{i_{t}}, a_{i_{t}})$ gives the period-$t$ utility of agent $i_{t}$, given his true preference $x_{i_{t}}$, report $\Tilde{x}_{i_{t}}$, and other agents' reports $\Tilde{x}_{-i_{t}}$.
Define the period-$t$ interim expected payoff of agent $i$, given $\alpha_{i}$, $<\sigma_{i}, \rho_{i}>$, and the interim process $\Gamma_{i,t}$, as follows
\begin{equation}\label{eq:interim_payoff_agent}
    \begin{split}
        C_{i_{t}} & (x_{i_{t}}, \tau_{i}; \alpha_{i}, \sigma_{i})
        \equiv \mathbb{E}\Big[\sum_{s=0}^{\tau_{i}-1} u_{i_{s}}\big(\hat{x}_{i_{s}}, \sigma_{i_{s}}(\alpha_{s}(\hat{\mathbf{x}}_{s} ))  \big) + \rho_{i_{s}}(\alpha_{s}(\hat{\mathbf{x}}_{s} )) \\
    &+ u_{i_{\tau_{i}}}\big(\hat{x}_{i_{\tau_{i}}}, \sigma_{i_{\tau_{i}}}(\alpha_{\tau_{i}}(\hat{\mathbf{x}}_{\tau_{i}} ))  \big) + \gamma_{i_{\tau_{i}}}(\alpha_{\tau_{i}}(\hat{\mathbf{x}}_{\tau_{i}}) ) + \beta_{i_{\tau_{i}}}(\tau_{i}) \big| \alpha_{i}, \sigma_{i}\Big], 
    \end{split}
\end{equation}
%
where $\hat{\mathbf{x}}_{s}=\{\hat{x}_{i_{s}}, \{\hat{x}_{e_{s}}\}_{e_{s}\in I_{s}\backslash{\{i_{s}\}}}\}$.
Similarly, we can define the ex-ante expected payoff of agent $i$, given $\alpha_{i}$, $<\sigma_{i}, \rho_{i}, \gamma_{i}, \beta>$, and the stochastic process $\Gamma_{i}$, as
\begin{equation}\label{eq:ex_ante_agent}
   \begin{split}
        C_{i}(&\tau_{i};\alpha_{i}, \sigma_{i}) 
        \equiv \mathbb{E}\Big[\sum_{s=0}^{\tau_{i}-1} u_{i_{s}}\big(\hat{x}_{i_{s}}, \sigma_{i_{s}}(\alpha_{s}(\hat{\mathbf{x}}_{s} ))  \big) + \rho_{i_{s}}(\alpha_{s}(\hat{\mathbf{x}}_{s} ))\\
        &+  u_{i_{\tau_{i}}}\big(\hat{x}_{i_{\tau_{i}}}, \sigma_{i_{\tau_{i}}}(\alpha_{\tau_{i}}(\hat{\mathbf{x}}_{\tau_{i}} ))  \big) + \gamma_{i_{\tau_{i}}}(\alpha_{\tau_{i}}(\hat{\mathbf{x}}_{\tau_{i}}) ) + \beta_{i_{\tau_{i}}}(\tau_{i}) \big| \alpha_{i}, \sigma_{i}\Big].
   \end{split}
\end{equation}

The reasoning process for the agents induced by the mechanism and the dynamic environment can be characterized by an incomplete-information game.
In this game, an agent only observes his preference by himself and reports it only to the CP.
Each agent $i$'s period-$t$ knowledge about other agents' preference is based on his realized allocations $a^{t}_{i}$ and the priors about the stochastic of others' preferences, i.e., $\mathbf{F}_{-i,0}$ and $\mathbf{K}_{-i}$.
%
%
In this work, we consider a direct mechanism (e.g., \cite{myerson1981optimal}), in which each agent $i$ finds it is his best interest to report his preference truthfully at each period $t$, i.e., $x_{i_{t}} = \alpha_{i_{t}}(x_{i_{t}})$, for all $t\in\mathbb{T}$.
The solution concept for the game is perfect Bayesian Nash equilibrium (PBNE), in which each agent reports truthfully at each period given the belief that all other agents report truthfully.
Let $b_{i_{t}}:X_{i_{t-1}} \times A_{i_{t-1}} \mapsto \Delta(\mathbf{X}_{-i_{t-1}})$ denote agent $i_{t}$'s belief about other agents' realized preferences $\mathbf{x}_{-i_{t-1}}$ given $ \tilde{x}_{i_{t-1}}$, and $a_{i_{t-1}}$, with $\mu_{i_{t}}(\cdot|\tilde{x}_{i_{t-1}},  a_{i_{t-1}})$ as the corresponding density function.
%
%
Let $\alpha^{*}_{i} = \{\alpha^{*}_{i_{t}}\}$ be agent $i$'s truthful reporting strategy, i.e., $\alpha^{*}_{i_{t}}(x_{i_{t}}) = x_{i_{t}}$, for all $x_{i_{t}}\in X_{i_{t}}$, $t\in\mathbb{T}$.
PBNE is defined as follows.

\begin{definition}\label{def:PBNE}
(PBNE) Given the mechanism $< \bm{\sigma}, \bm{\rho}, \bm{\gamma}, \bm{\beta}>$, the game of the agents admits a PBNE if, for all $x_{i_{t}}\in X_{i_{t}}$, $\tau_{i_{t}}\in \mathbb{T}_{t}$, $t\in\mathbb{T}$
\begin{equation}\label{eq:PBNE_optimal}
    \alpha^{*}_{i}=\arg_{\alpha_{i}}\max C_{i_{t}}(x_{i_{t}}, \tau_{i_{t}}; \alpha_{i},\sigma_{i}),
\end{equation}
and the belief $b_{i_{t}}$ is updated according Bayes' rule by receiving $<x_{i_{t}}, a_{i_{t}}>$, i.e.,
\begin{equation}\label{eq:PBNE_belief}
    \begin{split}
        \mu_{i_{t+1}}&(\mathbf{x}_{-i_{t+1}} | x_{i_{t}}, a_{i_{t}}) \\
        = &\frac{ \mu_{i_{t}}(\mathbf{x}_{-i_{t-1}}| x_{i_{t-1}}, a_{i_{t-1}}) \mathbf{f}_{-i}(\mathbf{x}_{-i_{t}}|\mathbf{x}_{-i_{t-1}}, \mathbf{a}_{i_{t-1}}) }{  \sum\limits_{\small{\substack{\mathbf{x}'_{-i_{t-1}}\in \mathbf{X}_{-i_{t-1}},\\ \mathbf{x}'_{-i_{t}}\in \mathbf{X}_{-i_{t}}} }} \mu_{i_{t}}(\mathbf{x}'_{-i_{t-1}}| x_{i_{t-1}}, a_{i_{t-1}}) \mathbf{f}_{-i}(\mathbf{x}'_{-i_{t}}|\mathbf{x}_{-i_{t-1}}, \mathbf{a}_{i_{t-1}})},
    \end{split}
\end{equation}
where $\mathbf{x}_{-i_{t-1}}$ and $\mathbf{a}_{i_{t-1}}$ in the update of $b_{i_{t}}$ are the estimated by agent $i$.
When the denominator of (\ref{eq:PBNE_belief}) become $0$, $\mu_{i_{t+1}}(\mathbf{x}_{-i_{t+1}} | x_{i_{t}}, a_{i_{t}})$ is updated by assigning any possible value. 
\end{definition}

Since each agent's preference is independently distributed within each period, agent $i$ can have a belief about each other individual agent's preference based on Bayes' rule in the similar way.
%
%

\subsection{Optimal Stopping}

Fix a truthful $\alpha^{*}_{i}$. For the convenience of notations, we omit the dependence on the reporting strategy when an agent reports truthfully, e.g., $C_{i}(\tau_{i};\sigma_{i}) = C_{i}(\tau_{i};\alpha^{*}_{i},\sigma_{i})$, unless otherwise stated.
Agent $i$'s stopping rule $\zeta_{i}$ is optimal if there exists a $\tau^{*}_{i}$ such that, for a given $<\sigma_{i}, \rho_{i}, \gamma_{i}, \beta_{i}>$, 
\begin{equation}\label{eq:optimal_stopping_definition}
    \sup_{\tau\in \mathbb{T}}C_{i}(\tau_{i};\alpha_{i}, \sigma_{i}) = C_{i}(\tau^{*}_{i};\alpha_{i}, \sigma_{i}).
\end{equation}
Define
$$
V_{i_{t}}(x_{i_{t}};\sigma_{i})\equiv \sup_{\tau_{i}\in \mathbb{T}_{t}} C_{i_{t}}(\tau_{i}; \sigma_{i}).
$$

\begin{assumption}\label{assp:stopping_assp}
For all $x_{i_{t}}$ $\in$ $X_{i_{t}}$, $t$ $\in$ $\mathbb{T}$, 
$\mathbb{E}\Big[\sup\limits_{\tau_{i}\in \mathbb{T}_{t}} \big|C_{i_{t}}(\tau_{i}; \sigma_{i})\big|  \Big]$ $<\infty.$
\end{assumption}

Suppose Assumption \ref{assp:stopping_assp} holds. Backward induction yields,
\begin{equation}\label{eq:Bellman}
    \begin{split}
        V_{i_{t}}(x_{i_{t}};\sigma_{i})=\max \Big(  C_{i_{t}}(x_{i_{t}}, t; \sigma_{i}),\;\; \mathbb{E}\Big[V_{i_{t+1}}(\hat{x}_{i_{t+1}};\sigma_{i}) \big| \sigma_{i} \Big]\Big), 
    \end{split}
\end{equation}
with $V_{i_{T}}(x_{i_{T}};\sigma_{i}) = C_{i_{T}}(T; \sigma_{i})$.
Let the stopping rule $\zeta_{i_{t}}$ be defined as
\begin{equation}\label{eq:optimal_stopping_rule}
\begin{split}
    \zeta_{i_{t}}: &\exists z\in \mathbb{T}, \text{ s.t., } \\
    z&=\inf\{t\in\mathbb{T}:  V_{i_{t}}(x_{i_{T}};\sigma_{i}) = C_{i_{t}}(x_{i_{t}},t; \sigma_{i}), x_{i_{t}}\in X_{i_{t}}\}. 
\end{split}
\end{equation}
It has been shown that the stopping rule (\ref{eq:optimal_stopping_rule}) is optimal (see, e.g., Theorem 1.9 of \cite{peskir2006optimal}), i.e., it solves problem (\ref{eq:optimal_stopping_definition}).
By $\zeta^{*}_{i}$, we denote the optimal stopping rule given in (\ref{eq:optimal_stopping_rule}).
%

\subsection{PBNE Honestly Implementable}

To successfully elicit agents' honest behaviors that admit a PBNE, we require an honesty elicitation (HE) constraint that (1) guarantees incentive compatibility and (2) maintains optimality of stopping rule.
We define the implementability of our mechanism in the following definition.
%
\begin{definition}\label{def:PBNE_HI}
(PBNE-HI) The mechanism $<\bm{\sigma}, \bm{\rho}, \bm{\gamma}, \bm{\beta}>$ satisfies PBNE-honesty implementable (PBNE-HI) if, for all reporting strategy $\alpha_{i}$, all $i_{t}\in I_{t}$, $t\in\mathbb{T}$, 
the following honesty elicitation (HE) constraint is satisfied:
\begin{equation}\label{eq:def_PBNE_HI}
    \begin{split}
        \max \Big( & C_{i_{t}}(x_{i_{t}}, t; \sigma_{i}),\;\; \mathbb{E}\Big[V_{i_{t+1}}(\hat{x}_{i_{t+1}};\sigma_{i}) \big| \sigma_{i} \Big]\Big)\\
        \geq&\max \Big(  C_{i_{t}}(x_{i_{t}}, t; \alpha_{i}, \sigma_{i}),\;\; \mathbb{E}\Big[V_{i_{t+1}}(\hat{x}_{i_{t+1}};\alpha_{i}, \sigma_{i}) \big|\alpha_{i}, \sigma_{i} \Big]\Big),
    \end{split}
\end{equation}
with $C_{i_{T}}(x_{i_{T}}, T; \sigma_{i})\geq C_{i_{T}}(x_{i_{T}}, T;\alpha_{i}, \sigma_{i})$,
and each agent believes that all other agents report truthfully all the time with probability $1$ and updates his belief according to (\ref{eq:PBNE_belief}).
\end{definition}

The condition (\ref{eq:def_PBNE_HI}) captures the coupling of each agent $i$' reporting decision and stopping decision. It elicits truthful reporting strategy $\alpha^{*}_{i}$ that is optimal at every period $t$ and the optimal stopping behavior according to $\zeta^{*}_{i}$.
That is, the condition (\ref{eq:def_PBNE_HI}) imposes that it is each agent $i$'s best interest to report truthfully at every period and make stopping decisions according to $\zeta^{*}_{i}$.

Given the mechanism rules, agent $i$'s decision makings at each period $t$ to maximizing $C_{i_{t}}$ requires the evaluation of future payoffs, which is determined by his current reporting strategy, his planned future reporting strategies, and the time horizon $\tau_{i_{t}}$. 
One the one hand, any $\{\alpha_{i_{s}}\}^{t}_{s}\subset \alpha_{i}$ determines (a set of) time horizon $\tau'_{i_{t}}\in\mathbb{T}_{t}$ such that $C_{i_{t}}(x_{i_{t}}, \tau'_{i_{t}};\alpha_{i}, \sigma_{i}) = \max_{\tau_{i}} C_{i_{t}}(x_{i_{t}}, \tau_{i_{t}};\alpha_{i}, \sigma_{i})$.
On the other hand, any time horizon $\tau_{i_{t}}\in\mathbb{T}_{t}$ can pin down (a set of) optimal reporting strategy $\{\alpha'_{i_{s}}\}^{t}_{s}\subset \alpha'_{i}$ that maximizes $C_{i_{t}}$, i.e., $C_{i_{t}}(x_{i_{t}}, \tau_{i_{t}};\alpha'_{i}, \sigma_{i}) = \max_{\alpha_{i}} C_{i_{t}}(x_{i_{t}}, \tau_{i_{t}};\alpha_{i}, \sigma_{i})$.
Additionally, agent $i$'s current reporting strategy $\alpha_{i_{t}}$ directly determines the immediate payoff and also indirectly affects the distribution of his future preferences which influence the expected future payoffs through the given mechanism rules.
Hence, condition (\ref{eq:def_PBNE_HI}) requires multi-dimensional guarantees. 

Let $\tilde{\alpha}_{i,(t)}=\{\tilde{\alpha}_{i_{s}}\}_{s\in\mathbb{T}}$ be any reporting strategy that differs from $\alpha^{*}_{i}$ only at period-$t$, (see it as one-shot deviation reporting strategy at period $t$) i.e., $\tilde{\alpha}_{i_{s}}=\alpha^{*}_{i_{s}}$, for all $s\in\mathbb{T}\backslash{\{t\}}$.
To emphasize the one shot deviation, in the rest of this letter we use $\tilde{x}_{i_{t}} = \tilde{\alpha}_{i_{t}}(x_{t})$ to denote the adoption of $\tilde{\alpha}_{i,(t)}$ unless otherwise stated.
We simplify the condition (\ref{eq:def_PBNE_HI}) by establishing a one-shot deviation principle.

\begin{proposition}\label{prop:one_shot_deviation}
The mechanism $<\bm{\sigma}, \bm{\rho},\bm{\gamma}, \bm{\beta}>$ is PBNE-HI if and only if there is no one-shot reporting strategy $\tilde{\alpha}_{i,(t)}$ with $\tilde{x}_{i_{t}} = \tilde{\alpha}_{i_{t}}(x_{t})$ at any period $t\in\mathbb{T}$, $i_{t}\in I_{t}$,
\begin{equation}\label{eq:def_PBNE_HI_one_shot}
    \begin{split}
        \max \Big( & C_{i_{t}}(x_{i_{t}}, t; \sigma_{i}),\;\; \mathbb{E}\Big[V_{i_{t+1}}(\hat{x}_{i_{t+1}};\sigma_{i}) \big| \sigma_{i} \Big]\Big)\\
        \geq&\max \Big(  C_{i_{t}}(x_{i_{t}}, t; \tilde{x}_{i_{t}}, \sigma_{i}),\;\; \mathbb{E}\Big[V_{i_{t+1}}(\hat{x}_{i_{t+1}};\tilde{x}_{i_{t}}, \sigma_{i}) \big|\tilde{x}_{i_{t}}, \sigma_{i} \Big]\Big),
    \end{split}
\end{equation}
with $C_{i_{T}}(x_{i_{T}}, T; \sigma_{i})\geq C_{i_{T}}(x_{i_{T}}, T;\tilde{x}_{i_{T}}, \sigma_{i})$.
\end{proposition}
\proof{}
See Appendix \ref{app:prop_one_shot_deviation}.
\endproof

At each period $t$, agent $i$'s reasoning about choosing stopping time characterized by a stopping rule is independent of the reporting strategy; however, the stopping decision according to $\zeta^{*}_{i}$ is coupled with the current reporting strategy.
Let 
%
{\setlength\abovedisplayskip{0.5pt}
\setlength\belowdisplayskip{0.5pt}
\begin{equation}\label{eq:period_t_max_future_time}
\setlength{\abovedisplayskip}{1pt}
\setlength{\belowdisplayskip}{1pt}  
    \tau^{*}_{i_{t}}(x_{i_{t}};\alpha_{i_{t}}) = \inf\{\tau\in \mathbb{T}: \tau\in\arg_{\tau'\in\mathbb{T}_{t}}C_{i_{t}}(\tau'; \alpha_{i_{t}}, \sigma_{i})\}.
\end{equation}}
%
%
Clearly, when $\tau^{*}_{i_{t}}(x_{i_{t}};\alpha_{i_{t}})=t$, it is optimal for agent $i$ to stop at $t$ given his true preference $x_{i_{t}}$ and period-$t$ reporting strategy $\alpha_{i_{t}}$.
Hence, for a given stopping rule, agent $i$'s any reporting strategy $\alpha'_{i_{t}}$ has an associated $\tau^{*}_{i_{t}}(x_{i_{t}};\alpha'_{i_{t}})$ for each observed preference $x_{i_{t}}$ and the optimal reporting strategy $\alpha''_{i_{t}}$ is given as,
$
\alpha''_{i_{t}}=\arg_{\alpha_{i_{t}}}\max C_{i_{t}}(\tau^{*}_{i_{t}}(x_{i_{t}};\alpha_{i_{t}}) ; \alpha_{i_{t}}, \sigma_{i}).
$
HE constraint elicits the truthful reporting $\alpha^{*}_{i}$ and optimal stopping rule $\zeta^{*}_{i}$ as the optimal decision rules for each agent $i$ in any PBNE-HI mechanism.
The analysis in the rest of this letter will focus on one-shot deviation report strategies.

\subsection{Threshold Rule}
Next, we introduce an important class of stopping rules that is based on a threshold rule (see, e.g., \cite{villeneuve2007threshold,kruse2018inverse}).
Define, for all $\tau_{i}\in \mathbb{T}_{t+1}$,
\begin{equation}\label{eq:continuing_value}
    \begin{split}
        &g_{i_{t}}(x_{i_{t}},\tau_{i};\tilde{x}_{i_{t}}, \sigma_{i})\\
        \equiv&\mathbb{E}\Big[ C_{i_{t+1}}(\hat{x}_{i_{t+1}}\tau_{i};\tilde{x}_{i_{t}},\sigma_{i}  )  | \sigma_{i}\Big] -C_{i_{t}}(x_{i_{t}}, t;\tilde{x}_{i_{t}}, \sigma_{i}) +\beta_{i_{t}}(t),
    \end{split}
\end{equation}
with $g_{i_{t}}(x_{i_{t}};\tilde{x}_{i_{t}}, \sigma_{i})= g_{i_{t}}(x_{i_{t}}, \tau'_{i};\tilde{x}_{i_{t}}, \sigma_{i})$, where $\tau'_{i}=\arg_{\tau_{i}\in \mathbb{T}_{t+1}}\sup\mathbb{E}\Big[ C_{i_{t+1}}(\hat{x}_{i_{t+1}}\tau_{i};\sigma_{i}  )  | \sigma_{i}\Big]$.
%

From the definition of the optimal stopping rule $\zeta^{*}_{i}$, we can define a stopping region as follows:
\begin{equation}\label{eq:stopping_region_0}
    D_{i_{t}}(t) \equiv \{x_{i_{t}}\in X_{i_{t}}: g_{i_{t}}(x_{i_{t}};\sigma_{i})\leq \beta_{i_{t}}(t))\}.
\end{equation}
We consider the following assumptions.

\begin{assumption}\label{assp:single_crossing} 
The difference of expected payoff between stopping at the next period $t+1$ and at the current period $t$,
$
C_{i_{t}}(x_{i_{t}}, t+1;\sigma_{i})-C_{i_{t}}(x_{i_{t}}, t;\sigma_{i})
$,
is non-decreasing in $x_{i_{t}}$, for all $t\in\mathbb{T}$.
\end{assumption}






Let $\epsilon_{i}:\mathbb{T} \mapsto X_{i_{t}}$ be the threshold function, such that agent $i_{t}$ decides to stop at period $t$ if $x_{i_{t}}\leq \epsilon_{i}(t)$.
Since all agents stop at the final period $T$, we make $\epsilon_{i}(T)=\bar{x}_{i_{t}}$.
The optimal threshold (stopping) rule is defined as follows:
\begin{equation}\label{eq:optimal_threshold_rule}
\setlength{\abovedisplayskip}{0.5pt}
\setlength{\belowdisplayskip}{1pt}
    \begin{split}
        \zeta^{\epsilon}_{i}: &\exists z\in \mathbb{T}, \text{ s.t., } z=\inf\{t\in\mathbb{T}:  x_{i_{t}}\leq \epsilon_{i}(t)\}. 
    \end{split}
\end{equation}

We establish the following proposition.
\begin{proposition}\label{prop:uniqueness_threshold}
Suppose Assumption \ref{assp:stochastic_dominance} holds.
In PBNE-HI mechanism $<\bm{\sigma}, \bm{\rho}, \bm{\gamma}, \bm{\beta}>$, each agent $i$'s optimal stopping rule is a threshold rule, i.e., $\zeta_{i}=\zeta^{\epsilon}_{i}$ with a unique threshold function $\epsilon_{i}$, for all $i\in I$.
%
\end{proposition}
\proof{}
See Appendix \ref{app:prop_uniqueness_threshold}.

\endproof

Proposition \ref{prop:uniqueness_threshold} implies that the threshold function $\epsilon_{i}$ can be characterized by the posted-price rule $\beta_{i}$, and vice versa. The following lemma follows Proposition \ref{prop:uniqueness_threshold}.

\begin{lemma}\label{lemma:beta_and_g_relation}
Fix truthful $\alpha^{*}_{i}$. Given $\alpha_{i}, \rho_{i}$, and $\beta_{i}$, we have $\beta_{i_{t}}(t) = g_{i_{t}}(\epsilon_{i}(t); \sigma_{i}).$
%
%
\end{lemma}

Lemma \ref{lemma:beta_and_g_relation} is based on the monotonicity of $g_{i_{t}}$ and the stopping region defined in (\ref{eq:stopping_region_0}).
\setlength{\belowdisplayskip}{0.5pt}

\subsection{CP's Problem}

The CP's design problem requires the estimation of population dynamics.
Without loss of generality, we assume the indices in $I_{t}$ are ordered according to the magnitude of $\tau^{*}_{e_{t}}(x_{e_{t}})$ given in (\ref{eq:period_t_max_future_time}).
Specifically, let $e_{t}$, $e'_{t}\in I_{t}$. Then, $e_{t}< e'_{t}$ if $\tau^{*}_{e_{t}}(x_{e_{t}})>\tau^{*}_{e_{t}}(x_{e'_{t}})$. When $\tau^{*}_{e_{t}}(x_{e_{t}})=\tau^{*}_{e_{t}}(x_{e'_{t}})$, the relative order is determined according to the order in the previous period.
For simplicity, we introduce the \textit{index mapping}  $\pi_{t}:I_{t-1}\mapsto I_{t}$ such that $i_{t}=\pi_{t}(i_{t-1})$ based on the aforementioned ordering rules. If agent $e_{t}$ leaves at $t$, then $\pi_{t}(e_{t})\not\in I_{t+1}$.
Let $\bm{\tau}^{*}=\{\tau^{*}_{i}\}_{i\in I}$ denote the \textit{mean first passage time} or \textit{expected stopping time} evaluated at the ex-ante stage.
The calculation of $\bm{\tau}^{*}=\{\tau^{*}_{i}\}_{i\in I}$ is beyond the scope of this letter and interested reader may refer to e.g., \cite{jaskowski2015first}.
Let $I^{S}_{t}\equiv\{i_{t}\in I_{t}: \pi_{t+1}(i_{t})\not\in I_{t+1}\}$ (i.e., set of period-$t$ leaving agents), $I^{\bar{S}}_{t}\equiv I_{t}\backslash I^{S}_{t}$, and $\mathbb{T}^{\backslash \bm{\tau}} \equiv \mathbb{T}_{\max{\bm{\tau}}}\backslash\bm{\tau}$ for any set $\bm{\tau}\in \mathbb{T}^{n_{t}}$.
Furthermore, denote $\mathbf{x}^{S}_{t}\equiv \{x_{i_{t}} \}_{i_{t}\in I^{S}_{t}}$ and $\mathbf{x}^{\bar{S}}_{t}= \mathbf{x}_{t}\backslash \mathbf{x}^{S}_{t}$.
The CP's design problem is to maximize the following constrained expected payoff at ex-ante by determining $<\bm{\sigma}, \bm{\rho}, \bm{\gamma},\bm{\beta}>$:
{\small
\begin{equation}\label{eq:CP_objective}
    \begin{split}
        \max_{\bm{\sigma}, \bm{\rho}, \bm{\gamma},\bm{\beta}}& C_{0}=\mathbb{E}\Big[\sum_{t\in \mathbb{T}^{\backslash\bm{\tau}^{*}}}u_{0,t}\big(\hat{\bm{x}}_{t},\bm{\sigma}_{t}(\hat{\bm{x}}_{t}) - \bm{\rho}_{t}(\hat{\bm{x}}_{t})  \big) \\
       &-\sum_{\tau\in \bm{\tau}^{*}}\Big(\sum_{e_{\tau}\in I^{\bar{S}}_{\tau}}\rho_{e_{t}}(\hat{x}_{e_{t}}) + \sum_{e_{\tau}\in I^{\bar{S}}_{\tau}}\big(\gamma_{e_{\tau}}(\hat{\bm{x}}_{\tau}) +  \beta_{e_{\tau}}(\tau) \big)  \Big) \Big],\\
       \text{s.t. }& \textit{PBNE-HI}\\
       & \text{ (PC) } C_{i}(\tau_{i};\sigma)\geq 0, \forall i\in I,
    \end{split}
\end{equation}
}
where (PC), which stands for participation constraint, incentivizes agents to participate.
In the following section, we characterize the PBNE-HI and provide design principle for honesty elicitation rules.


\section{Characterizing the Honesty Elicitation}
%
In this section, we characterize the honesty elicitation constraint of PBNE-HI given in (\ref{eq:def_PBNE_HI_one_shot}).
First, we introduce the notion of \textit{distance function}, defined as follows: for any $x_{i_{t}}$, $y_{i_{t}}\in X_{i_{t}}$, $\mathbf{x}_{-i_{t}}\in \mathbf{X}_{-i_{t}}$, $i_{t}\in I_{t}$,  $t\in\mathbb{T}$, 
\begin{equation}\label{eq:distance_stop}
    \begin{split}
        d^{S}_{i_{t}}(y_{i_{t}}, x_{i_{t}}; \sigma_{i})\equiv& u_{i_{t}}(y_{i_{t}}, \sigma_{i_{t}}(y_{i_{t}}, \mathbf{x}_{-i_{t}})) - u_{i_{t}}(x_{i_{t}}, \sigma_{i_{t}}(y_{i_{t}}, \mathbf{x}_{-i_{t}})),
    \end{split}
\end{equation}
and, for any $\tau_{i_{t}}\in \mathbb{T}_{t}$,
\begin{equation}\label{eq:distance_non_stop}
\setlength{\abovedisplayskip}{0.5pt}
\setlength{\belowdisplayskip}{0.5pt}
    \begin{split}
    &d^{\bar{S}}_{i_{t}}(y_{i_{t}}, x_{i_{t}}, \tau_{i_{t}};\sigma_{i_{t}})\equiv\max_{\tau\in \mathbb{T}_{\tau_{i_{t}}}}\beta_{\tau}(\tau) \\
        +&\mathbb{E}_{t}\Big[\sum_{s=0}^{\tau_{i_{t}}}u_{i_{s}}(\hat{x}_{i_{s}}, \sigma_{i_{s}}(\hat{x}_{i_{s}}, \hat{\mathbf{x}}_{-i_{s}})) +\gamma_{i_{ \tau_{i_{t}}  }}(\hat{x}_{i_{\tau_{i_{t}}}}, \hat{\bm{x}}_{-i_{\tau_{i_{t}}}})  | \sigma_{i}\Big]\Big|_{\hat{x}_{i_{t}}= y_{i_{t}}} \\
        -& \mathbb{E}_{t}\Big[\sum_{s=0}^{\tau_{i_{t}}}u_{i_{s}}(\hat{x}_{i_{s}}, \sigma_{i_{s}}(\hat{x}_{i_{s}}, \hat{\mathbf{x}}_{-i_{s}})) +\gamma_{i_{\tau_{i_{t}}}}(\hat{x}_{i_{\tau_{i_{t}}}}, \hat{\bm{x}}_{-i_{\tau_{i_{t}}}}) | y_{i_{t}},\sigma_{i}\Big]\Big|_{\hat{x}_{i_{t}} = x_{i_{t}}},
    \end{split}
\end{equation}
where the superscripts $S$ and $\bar{S}$ represent \textit{stop} and \textit{non-stop}, respectively.
Basically, the distance function $d^{S}_{i_{t}}(y_{i_{t}}, x_{i_{t}};\sigma_{i})$ (resp. $d^{\bar{S}}_{i_{t}}(y_{i_{t}}, x_{i_{t}; \tau_{i_{t}}}, \sigma_{i})$) represents the change in agent $i$'s expected utilities if he keeps reporting $y_{i_{t}}$ when his true preferences are $y_{i_{t}}$ and $x_{i_{t}}$, and he stops at period $t$ (resp. plan to stop at $\tau_{i_{t}}$ in the future).

Let $\phi^{S}_{i_{t}}(\cdot|\sigma_{i}): X_{i_{t}}\mapsto \mathbb{R}$ and $\phi^{\bar{S}}_{i_{t}}(\cdot| \sigma_{i}): X_{i_{t}}\mapsto \mathbb{R}$ be the \textit{characterizing functions} of agent $i$ that are modeled by the allocation rule $\sigma_{i}$.
Note that the input of $\phi^{S}_{i_{t}}$ and $\phi^{\bar{S}}_{i_{t}}$ is the agent's report.
Let $\phi^{y, z}_{i_{t}}(y_{i_{t}}, x_{i_{t}}|\sigma_{i}) = \phi^{y}_{i_{t}}(y_{i_{t}}|\sigma_{i}) - \phi^{z}_{i_{t}}(x_{i_{t}}|\sigma_{i})$, with $\phi^{y, z}_{i_{t}}( x_{i_{t}}|\sigma_{i}) = \phi^{y}_{i_{t}}(x_{i_{t}}|\sigma_{i}) - \phi^{z}_{i_{t}}(x_{i_{t}}|\sigma_{i})$, for $y,z\in\{S,\bar{S}\}$.
Construct the payment rules $<\rho_{i}, \gamma_{i}, \beta_{i}>$ in terms of the characterizing functions as follows:


%
\begin{equation}\label{eq:rho_characterization}
    \begin{split}
        \rho_{i_{t}}(x_{i_{t}}, \mathbf{x}_{-i_{t}})=& \phi^{\bar{S}}_{i_{t}}(x_{i_{t}}|\sigma_{i}) - \mathbb{E}_{t}\Big[ \phi^{\bar{S}}_{i_{t+1}}(\hat{x}_{i_{t+1}}|\sigma_{i})   |\sigma_{i}\Big] \\
        &- u_{i_{t}}(x_{i_{t}}, \sigma_{i_{t}}(x_{i_{t}}, \mathbf{x}_{-i_{t}})),
    \end{split}
\end{equation}
\begin{equation}\label{eq:gamma_characterization}
    \begin{split}
        \gamma_{i_{t}}(x_{i_{t}}, \mathbf{x}_{-i_{t}})
        =\phi^{S}_{i_{t}}(x_{i_{t}}|\sigma_{i}) -u_{i_{t}}(x_{i_{t}}, \sigma_{i_{t}}(x_{i_{t}}, \mathbf{x}_{-i_{t}})).
    \end{split}
\end{equation}
\begin{equation}\label{eq:beta_characterization}
    \begin{split}
        \beta_{i_{t}}(t) = \mathbb{E}\Big[ \sum_{s=t}^{T-1}\phi^{S,\bar{S}}_{i_{s+1}}(&\hat{x}_{i_{s+1}}\vee \epsilon_{i}(s+1)) \\
        &-  \phi^{S,\bar{S}}_{i_{s}}(\hat{x}_{i_{s}}\vee \epsilon_{i}(s))  \Big| x_{i_{t}} =  \epsilon_{i}(t),\sigma_{i}\Big].
    \end{split}
\end{equation}

We obtain a sufficient condition for PBNE-HI in the following proposition.

\begin{proposition}\label{prop:PBNE-HI_sufficient}
%
%
The mechanism $<\bm{\sigma}, \bm{\rho}, \bm{\gamma}, \bm{\beta}>$ is PBNE-HI if $\rho_{i_{t}}$ and $\gamma_{i_{t}}$ are constructed in (\ref{eq:rho_characterization}) and (\ref{eq:gamma_characterization}), respectively, and the characterizing functions satisfying the following conditions: for all $x_{i_{t}}$, $y_{i_{t}}\in X_{i_{t}}$, $i_{t}\in I_{t}$, and $t\in \mathbb{T}$,
\begin{equation}\label{eq:sufficient_condition_1}
    d^{S}_{i_{t}}(y_{i_{t}}, x_{i_{t}}|\sigma_{i}) \geq \phi^{S, S}_{i_{t}}(y_{i_{t}}, x_{i_{t}}|\sigma_{i}), 
\end{equation}
\begin{equation}\label{eq:sufficient_condition_2}
    \begin{split}
        \sup_{\tau_{i_{t}}\in \mathbb{T}_{t+1}}\Big\{ d^{\bar{S}}_{i_{t}}(y_{i_{t}}, x_{i_{t}}, \tau_{i};\sigma_{i})\Big\}
    \geq& \phi^{\bar{S}, \bar{S}}_{i_{t}}(y_{i_{t}}, x_{i_{t}}|\sigma_{i}),
    \end{split}
\end{equation}

\begin{equation}\label{eq:sufficient_condition_3}
\phi^{\bar{S}, S}_{i_{t}}(x_{i_{t}}|\sigma_{i}) \geq 0, \text{ with } \phi^{\bar{S}, S}_{i_{T}}(x_{i_{T}}|\sigma_{i}) = 0.
\end{equation}
%
%
\end{proposition}
%
\proof{}
See Appendix \ref{app:prop_PBNE-HI_sufficient}.

\endproof

%
%

The construction (\ref{eq:rho_characterization}) implies that the period-$t$ non-stopping instantaneous payoff is set to be the difference between the period-$t$ $\phi^{\bar{S}}_{i_{t}}$ and the expected period-$t+1$ $\phi^{\bar{S}}_{i_{t+1}}$ when the $x_{i_{t}}$ is reported at $t$.
The construction (\ref{eq:gamma_characterization}) implies that the period-$t$ stopping instantaneous payoff is set to be the period-$t$ $\phi^{S}_{i_{t}}$.
The posted-price rule $\beta_{i_{t}}$ is constructed in (\ref{eq:beta_characterization}) is independent of realized preferences.
The explicit formulations of $\phi^{S}_{i_{t}}$ and $\phi^{\bar{S}}_{i_{t}}$ can be pinned down by the necessity of PBNE-HI.
The following theorem establishes a necessary and sufficient condition for PBNE-HI.




Define an auxiliary function
\begin{equation}\label{eq:auxiliary}
    \begin{split}
        J_{i_{t}}(x_{i_{t}}, \tau_{i_{t}}; \sigma_{i})\equiv & u_{i_{t}}(x_{i_{t}}, \sigma_{i_{t}}(x_{i_{t}}, \mathbf{x}_{i_{t}})) + \gamma_{i_{t}}(x_{i_{t}}, \sigma_{i_{t}}(x_{i_{t}}, \mathbf{x}_{i_{t}}))\\
        &+ \beta_{i_{t}}(\tau_{i_{t}})\mathbf{1}_{\{\tau_{i_{t}} = t\}} + g_{i_{t}}(x_{i_{t}}, \tau_{i}; \sigma_{i}) \mathbf{1}_{\{\tau_{i_{t}} > t\}},
    \end{split}
\end{equation}
where $\mathbf{1}_{\{\cdot\}}$ is the indicator function.


%
%
We obtain a first-order condition for PBNE-HI in the following lemma.

\begin{lemma}\label{lemma:first_order}
In any PBNE-HI mechanism,
each $<\sigma_{i}, \rho_{i}, \gamma_{i}, \beta_{i}>$ satisfies 
{\small
\begin{equation}\label{eq:first_order_condiiton}
    \begin{split}
        \frac{\partial J_{i_{t}}(\ell,\tau_{i_{t}}; \sigma_{i}) }{\partial \ell}\Big|_{\ell = x_{i_{t}}}
        =& -\mathbb{E}_{t}\Big[\sum^{\tau_{i_{t}}}_{s=t}\frac{\partial u_{i_{s}}(\ell, \sigma_{i_{s}}( \hat{\mathbf{x}}_{s})  )  }{\partial \ell}\Big|_{\ell = \hat{x}_{i_{s}}}\\
        &\prod_{k=t+1}^{s}\frac{-\partial F_{i_{k}}(\hat{x}_{i_{k}}| \ell, \sigma_{i_{k-1}}(\hat{\bm{x}}_{k-1}))}{f_{i_{k}}(\hat{x}_{i_{k}}|\hat{x}_{i_{k-1}}, \sigma_{i_{k-1}}(\hat{\bm{x}}_{k-1}))\partial \ell }\Big|_{\ell = x_{i_{k-1}} }\Big| \sigma_{i}\Big].
    \end{split}
\end{equation}
}
%
%
\end{lemma}
\proof{}
See Appendix \ref{app:lemma_first_order}.

\endproof

The following theorem establishes a necessary and sufficient condition for PBNE-HI.
\begin{theorem}\label{thm:necessary_sufficient}
Any mechanism $<\bm{\sigma}, \bm{\rho}, \bm{\gamma}, \bm{\beta}>$ is PBNE-HI if and only if the $\rho_{i_{t}}$ and $\gamma_{i_{t}}$ are given in (\ref{eq:rho_characterization}) and (\ref{eq:gamma_characterization}), respectively, and the characterizing functions satisfy, for any arbitrary fixed $x^{\delta}_{i_{t}}\in X_{i_{t}}$, all $x_{i_{t}}\in X_{i_{t}}$, any $\epsilon_{i}$, $i_{t}\in I_{t}$, $t\in \mathbb{T}$,
\begin{equation}\label{eq:nece_sufficent_S}
    \begin{split}
        \phi^{S}_{i_{t}}(x_{i_{t}} )=\int^{x_{i_{t}}}_{x^{\delta}_{i_{t}}} h_{i_{t}}(t, \ell; \sigma_{i})d\ell,
    \end{split}
\end{equation}
and
\begin{equation}\label{eq:nece_sufficent_nonS}
    \begin{split}
        \phi^{\bar{S}}_{i_{t}}(x_{i_{t}})=\sup_{\tau_{i_{t}}\in \mathbb{T}_{t}}\int^{x_{i_{t}}}_{x^{\delta}_{i_{t}}} h_{i_{t}}(\tau_{i_{t}},\ell;\sigma_{i})d\ell,
    \end{split}
\end{equation}
where $h_{i_{t}}(\tau_{i_{t}}, x; \sigma_{i}) = \frac{\partial J_{i_{t}}(\ell,\tau_{i_{t}}; \sigma_{i}) }{\partial \ell}|_{\ell = x}$.
\end{theorem}
\proof{}
See Appendix \ref{app:thm_necessary_sufficient}.

\endproof

The condition in Theorem \ref{thm:necessary_sufficient} specifies the constructions of $\phi^{S}_{i}$ and $\phi^{\bar{S}}_{i}$ in terms of $\sigma_{i}$.
The term $h_{i_{t}}(\tau_{i_{t}}, x_{i_{t}}; \sigma_{i})$ captures the impact of marginal change of $x_{i_{t}}$ on the expected (future) payoff with horizon $\tau_{i_{t}}\in\mathbb{T}_{t}$, based on the necessity of PBNE-HI.
%
%
Here, $\phi^{\bar{S}}_{i_{t}}$ can be interpreted as the maximum advantage of the agent $i$ with period-$t$ true preference $x_{i_{t}}$ has over some arbitrary $x^{\delta}_{i_{t}}$, due to the CP's not knowing that agent $i$'s true $x_{i_{t}}$, when he looks into the future.
$\phi^{\bar{S}}_{i_{t}}$ can be interpreted analogously, When agent $i$ decides to stop at $t$.
%
%
If a $\sigma^{\star}_{i}$ with a fixed $x^{\delta}_{i_{t}}$ satisfies the necessary and sufficient conditions, constructing the payment rules based on (\ref{eq:rho_characterization})-(\ref{eq:beta_characterization}) leads to a PBNE-HI mechanism.
Furthermore, the same $\sigma^{\star}_{i}$ also satisfies the necessary and sufficient condition for any other fixed $\tilde{x}^{\delta}_{i_{t}}$ but with different payment rules due to the dependence of $\phi^{S}_{i_{t}}$ and $\phi^{\bar{S}}_{i_{t}}$ on $\tilde{x}^{\delta}_{i_{t}}$.
However, this non-uniqueness of payment rules does not impact the necessity and sufficiency of PBNE-HI because the conditions (\ref{eq:sufficient_condition_1})-(\ref{eq:sufficient_condition_3}) and (\ref{eq:nece_sufficent_S})-(\ref{eq:nece_sufficent_nonS}) can be reformulated to be independent of payments.

The construction (\ref{eq:beta_characterization}) implies that $\beta_{i_{t}}$ can be completely characterized by $\sigma_{i}$ and $\epsilon_{i}$, given $\Gamma_{i}$.
As a result, the design of $\beta_{i}$ can be equivalent to the design of $\alpha_{i}$ and $\epsilon_{i}$.
The following proposition describes the design principle for the relation between $\epsilon_{i}$ and $\beta_{i}$.

\begin{proposition}\label{prop:design_beta_principle}
Fix $\sigma_{i}$. Let $r_{i}=\{r_{i_{0}},\dots, r_{i_{T}}\}\in \mathbb{R}^{T+1}$ denote a sequence of payments.
Define
$$
\setlength{\abovedisplayskip}{1pt}
\setlength{\belowdisplayskip}{1pt}
\begin{aligned}
R_{i}[\sigma_{i},\epsilon_{i}]\equiv& \Big\{r_{i}\in\mathbb{R}^{T}:\mathbb{E}\Big[ r_{i_{t}} + \sum_{s=t}^{T-1}\big(\phi^{S,\bar{S}}_{i_{s}}(\hat{x}_{i_{s}}\vee \epsilon_{i}(s)) \\
&- \phi^{S,\bar{S}}_{i_{s+1}}(\hat{x}_{i_{s+1}}\vee \epsilon_{i}(s+1)) \big) \Big| x_{i_{t}} =  \epsilon_{i}(t),\sigma_{i}\Big] =0  \Big\}.
\end{aligned}
$$
The following statements are true.
\begin{itemize}
    \item[(1)] 
    The mechanism is pBNE-HI i.f.f.   $\{\beta_{i_{t}}(t)\}_{t\in\mathbb{T}}\in R_{i}[\sigma_{i},\epsilon_{i}]$, for any $\epsilon_{i_{t}}(t)\in X_{i_{t}}$, for all $i_{t}\in I_{t}$, $t\in \mathbb{T}$.
    \item[(2)] The mechanism is PBNE-HI without $\bm{\beta}$ i.f.f. there exists $\epsilon_{i}$ that solves the following equation:
    {\small
    $$
    \setlength{\abovedisplayskip}{1pt}
    \setlength{\belowdisplayskip}{1pt}
    \begin{aligned}
    \mathbb{E}&\Big[ \sum_{s=t}^{T-1}\big(\phi^{S,\bar{S}}_{i_{s}}(\hat{x}_{i_{s}}\vee \epsilon_{i}(s)) \\
&- \phi^{S,\bar{S}}_{i_{s+1}}(\hat{x}_{i_{s+1}}\vee \epsilon_{i}(s+1)) \big) \Big| x_{i_{t}} =  \epsilon_{i}(t),\sigma_{i}\Big] =0
    \end{aligned}
    $$}
    \item[(3)] Agent $i$ has no incentive to leave i.f.f.
    $\{g_{i_{t}}(\bar{x}_{i_{t}};\sigma_{i})\}_{t\in\mathbb{T}}$ $\in R_{i}[\sigma_{i},\bar{x}_{i_{t}}]$. 
    There is in general no global posted-price rule, i.e., $\beta_{i}=\beta$ for all $i\in I$, to disincentivize heterogeneous agents to leave until $T$, i.e., to prevent population dynamics.
\end{itemize}

\end{proposition}

\proof{}
See Appendix \ref{app:prop_design_beta_principle}.

\endproof
\subsection{Relaxation of CP's Problem}

To exactly solve the CP's mechanism design problem (\ref{eq:CP_objective}), the following criteria have to be satisfied: (1) PBNE-HI, (2) profit maximization, assuming agents' truthful reporting, and (3) computational constraints (e.g., scale, accuracy of approximations).
In general, however, the CP's mechanism design problem is analytically intractable.
Hence, approximations or proper relaxations are usually adopted to the aforementioned three criteria.
Since the scope of this work is to provide game-theoretic analysis for the implementability, we put the design of efficient algorithms to approximate the CP's problem in our future work and provide examples of relaxations of the game-theoretic constrains.




%

\begin{assumption}\label{assp:stochastic_dominance}
For any $x_{i_{t}}$, $x'_{i_{t}}\in X_{i_{t}}$ with $x_{i_{t}}\leq x'_{i_{t}}$, $\mathbf{x}_{-i_{t}}\in \mathbf{X}_{i_{t}}$, and $x_{i_{t+1}}\in X_{i_{t+1}}$, $t\in \mathbb{T}\backslash{T}$, the following holds:
{\small
$$
\setlength{\abovedisplayskip}{0.5pt}
\setlength{\belowdisplayskip}{0.5pt}
F_{i_{t+1}}(x_{i_{t+1}}|x_{t}, \sigma_{i_{t}}(x_{t}, \mathbf{x}_{-i_{t}})) \leq F_{i_{t+1}}(x_{i_{t+1}}|x'_{t}, \sigma_{i_{t}}(x'_{t}, \mathbf{x}_{-i_{t}})).
$$}
\end{assumption}

Let $\Pi_{t}: I\mapsto I_{t}$ such that $\Pi_{t}(i)= (\pi_{0}\circ\pi_{1}\cdots\circ \pi_{t})(i)=i_{t}$, for all $i\in I$, $t\in \mathbb{T}$, with $\Pi^{-1}_{t}(i_{t})=i$.

\begin{lemma}\label{lemma:relax_1}
Suppose $u_{i_{t}}$ is non-decreasing in $x_{i_{t}}$, for all $i_{t}\in I_{t}$, $t\in \mathbb{T}$ and Assumption \ref{assp:stochastic_dominance} holds.
Then, $\mathbb{E}\Big[ C_{i_{0}}(\underline{x}_{i_{0}}, \tau^{*}_{i_{0}};\sigma_{\Pi^{-1}_{0}(i_{0})})| \sigma_{\Pi^{-1}_{0}(i_{0})}\Big]\geq0$ implies the PC constraint in (\ref{eq:CP_objective}).
\end{lemma}

Based on Lemma \ref{lemma:relax_1}, we relax the CP's problem (\ref{eq:CP_objective}) by making $\mathbb{E}\Big[ C_{i_{0}}(\underline{x}_{i_{0}}, \tau^{*}_{i_{0}};\sigma_{\Pi^{-1}_{0}(i_{0})})| \sigma_{\Pi^{-1}_{0}(i_{0})}\Big] = 0$ and additionally making the CP to make decisions at a stationary point, i.e., the necessary conditions of PBNE-HI are satisfied.  
The relaxed problem is as follows:
{\small
\begin{equation}\label{eq:CP_reformulated_object}
\setlength{\abovedisplayskip}{1pt}
\setlength{\belowdisplayskip}{1pt}
    \begin{split}
        &\max_{\bm{\sigma}, \bm{\epsilon}}\;\;C_{0}= \mathbb{E}\Big[\sum\limits_{t\in\mathbb{T}^{\backslash \bm{\tau}^{*}}}u_{0,t}\big(\hat{\bm{x}}_{t},\bm{\sigma}_{t}(\hat{\bm{x}}_{t})  \big)\\
        + \sum_{i_{0}\in I, \tau_{i_{0}}\in \bm{\tau}^{*}} &\sum_{t=0, i_{t} =\Pi_{t}(i_{0}) }^{\tau_{i_{0}}}u_{i_{t}}(\hat{x}_{i_{t}}, \sigma_{i_{t}}(\hat{\bm{x}}_{t}))\\
        - \sum_{i_{0}\in I, \tau_{i_{0}}\in \bm{\tau}^{*}}&\sum_{t=0, i_{t} =\Pi_{t}(i_{0}) }^{\tau_{i_{0}}}\frac{\partial u_{i_{t}}(\ell, \sigma_{i_{s}}(\hat{\bm{x}}_{s})) (1-F_{i_{0}}(\hat{x}_{i_{0}} ))  }{\partial \ell f_{i_{0}}(\hat{x}_{i_{0}})(1/G_{i_{0}}(\hat{x}_{i_{0}} t))}  \Big|_{\ell = \hat{x}_{t}}
        %
        \Big| \bm{\sigma}\Big],\\
        \text{s.t. }& \textit{PBNE-HI}.
    \end{split}
\end{equation}
}

Since $\beta_{i_{t}}$ can be characterized by $\sigma_{i}$ and $\epsilon_{i}$ given $\Gamma_{i}$, the relationship between $\epsilon_{t}$ and $\tau_{i}$ is captured by the definition $\tau^{*}_{i}=\mathbb{E}\Big[\sum_{t=0}^{T}tF_{i_{t}}(\epsilon_{i}(t)|\tilde{\theta}_{i_{t-1}}, \sigma_{i_{t-1}}(\tilde{\theta}_{i_{t-1}}))\Big| \sigma_{i}\Big]$.
Let $\bm{\alpha}^{*}$ and $\bm{\epsilon}^{*}$ solve (\ref{eq:CP_reformulated_object}).
Then, $\rho_{i_{t}}$, $\gamma_{i_{t}}$, and $\beta_{i_{t}}$ can be constructed by (\ref{eq:rho_characterization})-(\ref{eq:beta_characterization}), respectively.
Analytical relaxations of DIC may be used when there are computational burdens to check the DIC conditions given specific mathematical descriptions of economic model. Analytical relaxations can be imposed by reducing the necessity or the sufficiency of PBNE-HI such that the relaxed conditions imply the original ones. For example, the following two conditions imply (\ref{eq:sufficient_condition_1})-(\ref{eq:sufficient_condition_2}) by reducing the necessity, respectively:
{\small
\begin{equation}\label{eq:relax_suffi_1}
\big[h_{i_{t}}(t,y_{i_{t}};\sigma_{i}) -h_{i_{t}}(t,x_{i_{t}};y_{i_{t}},\sigma_{i}) \big]\big[ y_{i_{t}} - x_{i_{t}}\big]\geq 0,
\end{equation}
\begin{equation}\label{eq:relax_suffi_2}
    \begin{aligned}
&\big[\sup_{\tau\in\mathbb{T}_{t+1}}h_{i_{t}}(\tau, x_{i_{t}};y_{i_{t}},\sigma_{i})-\sup_{\tau\in\mathbb{T}_{t}}h_{i_{t}}(\tau, y_{i_{t}};\sigma_{i})\big]\big[y_{i_{t}} - x_{i_{t}}\big]\\
\leq&  \sup_{\tau\in\mathbb{T}_{t+1}}\Big\{\max_{\tau'\in\mathbb{T}_{\tau}} \beta_{i_{\tau'}}(\tau')+ \big[h_{i_{t}}(\tau,y_{i_{t}};\sigma_{i}) -  h_{i_{t}}( x_{i_{t}};y_{i_{t}},\sigma_{i})\big] \big[y_{i_{t}} - x_{i_{t}}\big]\Big\}.
    \end{aligned} 
\end{equation}
}
Computational constraints and intractability of (\ref{eq:CP_reformulated_object}) may require approximation techniques to compute sub-optimal solutions, which may inevitably violate the PBNE-HI.
%
%
Typical computational relaxations of the incentive compatibility (IC) involves considering $\delta$-IC ($\delta_{t}$-DIC in our case) that requires no agent can improve his payoff more than $\delta \geq 0$ when he misreports his private information.
%
%
The conditions of PBNE-HI can be used to measure the (violation) of DIC of any computationally relaxed mechanisms, which is an active research agenda in algorithmic mechanism design.

\section{Conclusion}

In this letter, we have presented a dynamic mechanism design problem for resource allocation to multiple self-interested agents in a dynamic environment due to agent's learning-by-doing and the adoption of stopping rules.
The mechanism consists of an allocation rule and three payment rules. 
One of the payment rule that is independent of agents' private preference is used to directly influence the population dynamics.
%
%
We have defined the notion of honesty elicitation to describe the scenario where each agent is incentivized to truthfully reveal his true preference to the central planner and follow an optimal stopping rule.
Necessary and sufficient conditions are established for the implementability of honesty elicitation under the perfect Bayesian Nash equilibrium.
We have presented two approaches to relax the mechanism design problem by (1) modifying the central planner's expected payoff and (2) relaxing the necessity of PBNE-HI.

\appendix

Let $\hat{\bm{x}}_{-i_{t}}$ and $\bm{x}_{-i_{t}}$ denote the preference (random variable) and its period-$t$ realization of all agents other than agent $i_{t}$.

\subsection{Proof of Proposition \ref{prop:one_shot_deviation} }\label{app:prop_one_shot_deviation}

We omit the \textit{only if} part that can be proved straightforwardly by following the optimality of honesty.
We prove the \textit{if} part by contradiction.
Suppose that the truthful reporting strategy $\alpha^{*}_{i}$ satisfies (\ref{eq:def_PBNE_HI_one_shot}) for any one-shot deviation reporting strategy $\tilde{\alpha}_{i,(t)}$ at period $t$, but it does not satisfy (\ref{eq:def_PBNE_HI}).
This means that there exists another reporting strategy $\alpha^{1}_{i}=\{\alpha^{1}_{i_{t}}\}_{t\in \mathbb{T}}$ and a preference $x_{i_{t}}$ such that 
$V_{i_{t}}(x_{i_{t}};\alpha^{1}_{i}, \sigma_{i}) > V_{i_{t}}(x_{i_{t}}; \sigma_{i})$.
Let $\zeta^{*}_{i}[\alpha^{*}_{i}]$ and $\zeta^{*}_{i}[\alpha^{1}_{i}]$ be the optimal stopping rules with $\alpha^{*}_{i}$ and $\alpha^{1}_{i}$, respectively.
Suppose that $\zeta^{*}_{i}[\alpha^{*}_{i}]$ calls for stopping at period $t$ but $\zeta^{*}_{i}[\alpha^{1}_{i}]$ calls for continuing; i.e. 
$$
\begin{aligned}
C_{i_{t}}(x_{i_{t}},t;\sigma_{i})< \mathbb{E}\Big[V_{i_{t+1}}(\hat{x}_{i_{t+1}}; \alpha^{1}_{i},\sigma_{i})\Big| \alpha^{1}_{i},\sigma_{i} \Big].
\end{aligned}
$$
Then, we can find some $\eta>0$ such that
\begin{equation}\label{eq:app_A_1}
    \begin{aligned}
\mathbb{E}\Big[V_{i_{t+1}}(\hat{x}_{i_{t+1}}; \alpha^{1}_{i},\sigma_{i})\Big| \alpha^{1}_{i},\sigma_{i} \Big]\geq C_{i_{t}}(x_{i_{t}},t;\sigma_{i}) +2\eta.
\end{aligned}
\end{equation}
%
Let $\alpha^{2}_{i}=\{\alpha^{2}_{i_{t}}\}_{t\in\mathbb{T}}$ be any reporting strategy.
Suppose that $\alpha^{2}_{i_{s}} = \alpha^{1}_{i_{s}}$, for all $s\in \mathbb{T}_{t,t+k}$, for some $k>0$ and
\begin{equation}\label{eq:app_A_2}
    \begin{aligned}
\mathbb{E}\Big[V_{i_{t+1}}(\hat{x}_{i_{t+1}}; \alpha^{2}_{i},\sigma_{i})&\Big| \alpha^{2}_{i},\sigma_{i} \Big] \\
\geq& \mathbb{E}\Big[V_{i_{t+1}}(\hat{x}_{i_{t+1}}; \alpha^{1}_{i},\sigma_{i})\Big| \alpha^{1}_{i},\sigma_{i} \Big] -\eta.
\end{aligned}
\end{equation}

Then, we have the following from (\ref{eq:app_A_1}) and (\ref{eq:app_A_2}):
\begin{equation}\label{eq:app_A_3}
    \begin{aligned}
    C_{i_{t}}(x_{i_{t}},t;\sigma_{i}) +\eta \leq \mathbb{E}\Big[V_{i_{t+1}}(\hat{x}_{i_{t+1}}; \alpha^{2}_{i},\sigma_{i})\Big| \alpha^{2}_{i},\sigma_{i} \Big]. 
    \end{aligned}
\end{equation}

Let $\bar{\alpha}_{i|t,t+k}=\{\bar{\alpha}_{i_{t}}\}_{t\in\mathbb{T}}$ be any reporting strategy such that $\bar{\alpha}_{i_{s}} = \alpha^{2}_{i_{s}}$, for all $s\in \mathbb{T}_{t,t+k}$ and $\bar{\alpha}_{i_{s}} = \alpha^{*}_{i_{s}}$ for all $s\in \mathbb{T}\backslash \mathbb{T}_{t,t+k}$, for some $k>0$.
From (\ref{eq:app_A_3}), we can see that using $\bar{\alpha}_{i|t,t+k}$ as a way to deviate from honesty is enough to gain profit.

Let $\alpha^{3}_{i|s}=\{\alpha^{3}_{i_{t}}\}_{t\in\mathbb{T}}$ be the one-shot deviation reporting strategy at period $s\in \mathbb{T}_{t,t+k}$, for $k>0$, such that $\alpha^{3}_{i_{s}} = \alpha^{2}_{i_{s}}$.
From (\ref{eq:app_A_3}), we have
\begin{equation}\label{eq:app_A_4}
    \begin{split}
        C_{i_{t}}(x_{i_{t}},t;\sigma_{i}) < \mathbb{E}\Big[V_{i_{t+1}}(\hat{x}_{i_{t+1}}; \alpha^{3}_{i|t+k-1},\sigma_{i})\Big| \alpha^{3}_{i|t+k-1},\sigma_{i} \Big].
    \end{split}
\end{equation}

Let $t'=t+k$.  
For all $x_{i_{t'}}\in X_{i_{t'-1}}$, we have
\begin{equation}\label{eq:app_A_5}
    \begin{split}
  V_{i_{t'-1}}(x_{i_{t'-1}};&\alpha^{3}_{i|t'-2},\sigma_{i}) \\
  =& \mathbb{E}\Big[V_{i_{t'-1}}(\hat{x}_{i_{t'-1}};  \alpha^{3}_{i|t'-2},\sigma_{i})\Big| \alpha^{3}_{i|t'-2},\sigma_{i}   \Big],    
  \end{split}
\end{equation}
and
\begin{equation}\label{eq:app_A_6}
    \begin{split}
        V_{i_{t'-1}}(x_{i_{t'-1}};\alpha^{3}_{i|t'-1},\sigma_{i})
        =& \max\Big(C_{i_{t'-1}}(x_{i_{t'-1}}, t'-1;\alpha^{3}_{i|t'-1}, \sigma_{i})  ,\\
        &\mathbb{E}\Big[V_{i_{t'}}(\hat{x}_{i_{t'}}; \alpha^{3}_{i|t'-1}, \sigma_{i})  \Big| \alpha^{3}_{i|t'-1}, \sigma_{i}\Big]\Big).
    \end{split}
\end{equation}
Then, since $\alpha^{*}_{i}$ satisfy (\ref{eq:def_PBNE_HI_one_shot}), we have
\begin{equation}\label{eq:app_A_7}
    \begin{split}
        V_{i_{t'-1}}(x_{i_{t'-1}}; \alpha^{3}_{i,t'-2},\sigma_{i})\geq V_{i_{t'-1}}(x_{i_{t'-1}}; \alpha^{3}_{i,t'-1},\sigma_{i})
    \end{split}
\end{equation}
Hence, (\ref{eq:app_A_5})-(\ref{eq:app_A_7}) imply
\begin{equation}\label{eq:app_A_8}
    \begin{split}
        \mathbb{E}\Big[V_{i_{t-1}}(\hat{x}_{i_{t-1}};  \alpha^{3}_{i|t'-2},\sigma_{i})&\Big| \alpha^{3}_{i|t'-2},\sigma_{i}   \Big]\\
        \geq & \mathbb{E}\Big[V_{i_{t-1}}(\hat{x}_{t-1};\alpha^{3}_{i_{t'-1}}, \sigma_{i})  \Big| \alpha^{3}_{i|t'-1}, \sigma_{i}\Big].
    \end{split}
\end{equation}
From (\ref{eq:app_A_4}), we have
\begin{equation}\label{eq:app_A_8}
    \begin{split}
        \mathbb{E}\Big[V_{i_{t-1}}(\hat{x}_{i_{t-1}};  \alpha^{3}_{i|t'-2},\sigma_{i})&\Big| \alpha^{3}_{i|t'-2},\sigma_{i}   \Big]> C_{i_{t}}(x_{i_{t}},t;\sigma_{i}).
    \end{split}
\end{equation}
Hence, we can use backward induction to obtain the following:
$$
\begin{aligned}
\mathbb{E}\Big[V_{i_{t-1}}(\hat{x}_{i_{t-1}};  \alpha^{3}_{i|t},\sigma_{i})&\Big| \alpha^{3}_{i|t},\sigma_{i}   \Big]> C_{i_{t}}(x_{i_{t}},t;\sigma_{i}),
\end{aligned}
$$
which contradicts the fact that $\alpha^{*}_{i}$ satisfies (\ref{eq:def_PBNE_HI_one_shot}).

We can use the similar procedures for the following cases: those are (1) $\zeta^{*}_{i}[\alpha^{*}_{i}]$ calls for stopping and $\zeta^{*}_{i}[\alpha^{1}_{i}]$ calls to stopping, (2) $\zeta^{*}_{i}[\alpha^{*}_{i}]$ calls for continuing and $\zeta^{*}_{i}[\alpha^{1}_{i}]$ calls to continuing, and (3) $\zeta^{*}_{i}[\alpha^{*}_{i}]$ calls for continuing and $\zeta^{*}_{i}[\alpha^{1}_{i}]$ calls to stop.
The one-shot deviation principle can then be proved.

\qed

\subsection{Proof of Proposition \ref{prop:uniqueness_threshold}}\label{app:prop_uniqueness_threshold}

Suppose there exist two threshold function $\epsilon'_{i}\neq \epsilon_{i}$. 
Let $\zeta^{\epsilon'}_{i}$ and $\zeta^{\epsilon}_{i}$ be the optimal stopping rule in PBNE-HI with respect to $\epsilon'_{i}$ and $\epsilon_{i}$, respectively.
As defined in (\ref{eq:period_t_max_future_time}), let $\tau^{\epsilon}_{i_{t}}(x_{i_{t}})$ and $\tau^{\epsilon'}_{i_{t}}(x_{i_{t}})$ be corresponding to $\zeta^{\epsilon}_{i}$ and $\zeta^{\epsilon'}_{i}$, respectively.
Suppose $\tau^{\epsilon'}_{i_{t}}(x_{i_{t}})=\tau^{\epsilon}_{i_{t}}(x_{i_{t}})$, for $x_{i_{t}}\in X_{i_{t}}$, $t\in \mathbb{T}$.
%
Assume without loss of generality $\epsilon_{i}(t)< \epsilon'_{i}(t)$ for some $t\in \mathbb{T}$.
%
%
Then, we have
$$
\begin{aligned}
P_{r}(\tau^{\epsilon}_{i_{t}}(\hat{x}_{i_{t}}) = t ) =& P_{r}(x_{i_{t}} \leq \epsilon_{i}(t), \tau^{\epsilon}_{i_{t-1}}(\hat{x}_{i_{t-1}})>t-1)  \\
=& \mathbb{E}\Big[\mathbb{E}\Big[\mathbf{1}_{\{\hat{x}_{i_{t}}\leq \epsilon_{i}(t) \}} \Big|\Gamma_{i_{t}}  \Big] \mathbf{1}_{\{ \hat{x}_{i_{t-1}}> \epsilon_{i}(t-1) \}} \Big| \Gamma_{i_{t-1}}  \Big],
\end{aligned}
$$
where the interim process in the expectation are used to indicate the how the expectation is the measured.
$P_{r}(\tau^{\epsilon'}_{i_{t}}(\hat{x}_{i_{t}}) = t )$ can be determined in the similar way.
Hence,
\begin{equation}\label{eq:app_B_7}
    \begin{aligned}
P_{r}(\tau^{\epsilon'}_{i_{t}}(&\hat{x}_{i_{t}}) = t )-P_{r}(\tau^{\epsilon}_{i_{t}}(\hat{x}_{i_{t}}) = t )\\
=& \mathbb{E}\Big[\mathbb{E}\Big[\mathbf{1}_{\{\hat{x}_{i_{t}}\leq \epsilon'_{i}(t) \}} \Big|\Gamma_{i_{t}}  \Big] \mathbf{1}_{\{ \hat{x}_{i_{t-1}}> \epsilon'_{i}(t-1) \}} \Big| \Gamma_{i_{t-1}}  \Big]\\
&-\mathbb{E}\Big[\mathbb{E}\Big[\mathbf{1}_{\{\hat{x}_{i_{t}}\leq \epsilon_{i}(t) \}} \Big|\Gamma_{i_{t}}  \Big] \mathbf{1}_{\{ \hat{x}_{i_{t-1}}> \epsilon_{i}(t-1) \}} \Big| \Gamma_{i_{t-1}}  \Big]\\
=&\mathbb{E}\Big[\mathbb{E}\Big[\mathbf{1}_{\{\epsilon_{i}(t)\leq \hat{x}_{i_{t}}\leq \epsilon'_{i}(t) \}} \Big|\Gamma_{i_{t}}  \Big] \mathbf{1}_{\{\tau^{\epsilon}_{i_{t-1}}(\hat{x}_{i_{t}})>t-1  \}} \Big| \Gamma_{i_{t-1}}  \Big].
\end{aligned}
\end{equation}
Since $\tau^{\epsilon'}_{i_{t}}(x_{i_{t}})=\tau^{\epsilon}_{i_{t}}(x_{i_{t}})$, $P_{r}(\tau^{\epsilon'}_{i_{t}}(\hat{x}_{i_{t}}) = t )-P_{r}(\tau^{\epsilon}_{i_{t}}(\hat{x}_{i_{t}}) = t )=0$.
However, since $f_{i_{t}}(x_{i_{t}})>0$ for all $x_{i_{t}}\in X_{i_{t}}$, $t\in\mathbb{T}$, the right hand side of (\ref{eq:app_B_7}) is non zero,
which provides a contradiction.

\qed

\subsection{Proof of Proposition \ref{prop:PBNE-HI_sufficient} }\label{app:prop_PBNE-HI_sufficient}

Define
\begin{equation}\label{eq:app_D_1}
    \begin{split}
        J^{S}_{i_{t}}(x_{i_{t}} ;\tilde{x}_{i_{t}},\sigma_{i}) = & u_{i_{t}}(x_{i_{t}}, \sigma_{i_{t}}(\tilde{x}_{i_{t}}, \hat{\bm{x}}_{-i_{t}}))+ \gamma_{i_{t}}(\tilde{x}_{i_{t}}, \bm{x}_{-i_{t}}) + \beta_{i_{t}}(t),
    \end{split}
\end{equation}
and
\begin{equation}\label{eq:app_D_2}
    \begin{split}
        J^{\bar{S}}_{i_{t}}(x_{i_{t}}; \tilde{x}_{i_{t}}, \sigma_{i}) =& u_{i_{t}}(x_{i_{t}}, \sigma_{i_{t}}(\tilde{x}_{i_{t}}, \hat{\bm{x}}_{-i_{t}}))+ \gamma_{i_{t}}(\tilde{x}_{i_{t}}, \bm{x}_{-i_{t}}) \\
        &+ g_{i_{t}}(x_{i_{t}};\tilde{x}_{i_{t}}, \sigma_{i}).
    \end{split}
\end{equation}
From the constructions of $\gamma_{i_{t}}$ in (\ref{eq:gamma_characterization}), we can represent $J^{S}_{i_{t}}$ in terms of the characterizing functions as follows:
\begin{equation}\label{eq:app_D_3}
    \begin{split}
        J^{S}_{i_{t}}(x_{i_{t}};\tilde{x}_{i_{t}},\sigma_{i}) = & u_{i_{t}}(x_{i_{t}}, \sigma_{i_{t}}(\tilde{x}_{i_{t}}, \bm{x}_{-i_{t}})) - u_{i_{t}}(\tilde{x}_{i_{t}}, \sigma_{i_{t}}(\tilde{x}_{i_{t}}, \bm{x}_{-i_{t}}))\\
        &+ \phi^{S}_{i_{t}}(\tilde{x}_{i_{t}}|\sigma_{i})+\beta_{i_{t}}(t).
    \end{split}
\end{equation}
From the definitions of the distance function $d^{S}_{i_{t}}$ in (\ref{eq:distance_stop}), we have
\begin{equation}\label{eq:app_D_5}
    \begin{aligned}
J^{S}_{i_{t}}(x_{i_{t}}; \tilde{x}_{i_{t}}, \sigma_{i})=& \phi^{S}_{i_{t}}(\tilde{x}_{i_{t}}|\sigma_{i}) + \beta_{i_{t}}(t) - d^{S}_{i_{t}}(\tilde{x}_{i_{t}}, x_{i_{t}}; \sigma_{i})\\
\leq& \phi^{S}_{i_{t}}(x_{i_{t}}|\sigma_{i}) + \beta_{i_{t}}(t) \text{\;\; From (\ref{eq:sufficient_condition_1})}\\
=& u_{i_{t}}(x_{i_{t}}, \sigma_{i_{t}}(x_{i_{t}}, \bm{x}_{i_{t}})) + \gamma_{i_{t}}(x_{i_{t}}, \bm{x}_{i_{t}}) + \beta_{i_{t}}(t)\\
=& J^{S}_{i_{t}}(x_{i_{t}}; x_{i_{t}}, \sigma_{i}).
\end{aligned}
\end{equation}

For $J^{\bar{S}}_{i_{t}}$, we first re-write it as follows:
\begin{equation}\label{eq:app_D_4}
    \begin{split}
        &J^{\bar{S}}_{i_{t}}(x_{i_{t}}; \tilde{x}_{i_{t}}, \sigma_{i}) \\
        =& u_{i_{t}}(x_{i_{t}}, \sigma_{i_{t}}(\tilde{x}_{i_{t}}, \bm{x}_{i_{t}})) +\rho_{i_{t}}(\tilde{x}_{i_{t}}, \bm{x}_{i_{t}}) ) -\beta_{i_{t}}(t)\\
        &+\sup_{\tau_{i_{t}}\in \mathbb{T}_{t+1}}\Big\{\mathbb{E}\Big[\phi^{\bar{S}}_{t+1}(\hat{x}_{i_{t+1}}|\sigma_{i}) - \phi^{\bar{S}}_{i_{\tau_{i_{t}}}}(\hat{x}_{i_{\tau_{i_{t}}}}) \\
        &+ \phi^{S}_{i_{\tau_{i_{t}}}}(\hat{x}_{i_{\tau_{i_{t}}}})+\beta_{i_{\tau_{i_{t}}}} ( \tau_{i_{t}})\Big| \tilde{x}_{i_{t}}, \sigma_{i}\Big]  \Big\}\\
        =& \sup_{\tau_{i}\in \mathbb{T}_{t+1}}\mathbb{E}\Big[\sum^{\tau_{i}}_{s=t} u_{i_{s}}(\hat{x}_{i_{s}}, \sigma_{i_{s}}(\hat{x}_{i_{s}}, \hat{\bm{x}}_{-i_{s}})) \\
        &+\sum_{s=t}^{\tau_{i}-1}\rho_{i_{s}}(\hat{x}_{i_{s}}, \hat{\bm{x}}_{-i_{s}}) \\
        &+ \gamma_{i_{\tau_{i}}} (\hat{x}_{i_{\tau_{i}}}, \hat{\bm{x}}_{-i_{\tau_{i}}}) +\beta_{i_{\tau_{i}}}(\tau_{i})  \Big| \tilde{x}_{i_{t}}, \sigma_{i} \Big]\Big|_{\hat{x}_{i_{t}} = x_{i_{t}}}.
    \end{split}
\end{equation}


The construction of $\rho_{i_{t}}$ in (\ref{eq:rho_characterization}) yields
\begin{equation}\label{eq:app_D_7}
    \begin{split}
        &\text{R.H.S of (\ref{eq:app_D_4})}\\
        =& \sup_{\tau_{i}\in \mathbb{T}_{t+1}}\Bigg\{ \mathbb{E}\Big[\sum_{s=t}^{\tau_{i}}u_{i_{s}}(\hat{x}_{t}, \sigma_{i_{s}}(\hat{x}_{t}, \hat{\bm{x}}_{-i_{s}}  )) \\
        &+ \gamma_{\tau_{i}}( \hat{x}_{\tau_{i} }, \hat{\bm{x}}_{-i_{\tau_{i} }} )| \tilde{x}_{i_{t}}, \sigma_{i}\Big]\Big|_{\hat{x}_{i_{t}} = x_{i_{t}}}+ \phi^{\bar{S}}_{i_{t}}(\tilde{x}_{i_{t}})\\
        &- \mathbb{E}\Big[ \phi^{\bar{S}}_{\tau_{i}}(\hat{x}_{\tau_{i}}) + \sum^{\tau_{i}-1}_{s=t}u_{i_{s}}(\hat{x}_{t}, \sigma_{i_{s}}(\hat{x}_{t}, \hat{\bm{x}}_{-i_{s}}  )) \Big| \sigma_{i}\Big]  \Big|_{\hat{x}_{i_{t}} = \tilde{x}_{i_{t}}} +\beta_{\tau_{i}}(\tau_{i}) \Bigg\}\\
        \leq&  \sup_{\tau_{i}\in \mathbb{T}_{t+1}}\Bigg\{ \mathbb{E}\Big[\sum_{s=t}^{\tau_{i}}u_{i_{s}}(\hat{x}_{t}, \sigma_{i_{s}}(\hat{x}_{t}, \hat{\bm{x}}_{-i_{s}}  )) \\
        &+ \gamma_{\tau_{i}}( \hat{x}_{\tau_{i} }, \hat{\bm{x}}_{-i_{\tau_{i} }} )| \tilde{x}_{i_{t}}, \sigma_{i}\Big]\Big|_{\hat{x}_{i_{t}} = x_{i_{t}}}+ \phi^{\bar{S}}_{i_{t}}(\tilde{x}_{i_{t}})\\
        &- \mathbb{E}\Big[ \phi^{\bar{S}}_{\tau_{i}}(\hat{x}_{\tau_{i}}) + \sum^{\tau_{i}-1}_{s=t}u_{i_{s}}(\hat{x}_{t}, \sigma_{i_{s}}(\hat{x}_{t}, \hat{\bm{x}}_{-i_{s}}  )) \Big| \sigma_{i}\Big]  \Big|_{\hat{x}_{i_{t}} = \tilde{x}_{i_{t}}} +\max_{\tau\in \mathbb{T}_{\tau_{i_{t}}}}\beta_{\tau}(\tau). \Bigg\},
    \end{split}
\end{equation}
%
%
From the definition (\ref{eq:distance_non_stop}) and condition (\ref{eq:sufficient_condition_3}), we have
\begin{equation}\label{eq:app_D_8}
    \begin{split}
        \text{R.H.S. of (\ref{eq:app_D_7})}\leq&\phi^{\bar{S}}_{i_{t}}(\tilde{x}_{i_{t}})-\sup_{\tau_{i}\in \mathbb{T}_{t+1}}\Bigg\{ d^{\bar{S}}_{i_{t}}(\tilde{x}_{t_{t}}, x_{t},\tau_{i};\sigma_{i}) \Bigg\} .
    \end{split}
\end{equation}
Then, condition (\ref{eq:sufficient_condition_2}) yields
\begin{equation}\label{eq:app_D_9}
    \begin{aligned}
    \text{R.H.S. of (\ref{eq:app_D_8})}\leq \phi^{\Bar{S}}_{i_{t}}(x_{i_{t}}).
    \end{aligned}
\end{equation}
From the construction of $\rho_{i_{t}}$ in (\ref{eq:rho_characterization}) and condition (\ref{eq:sufficient_condition_3}), we have, for any $\tau_{i}\in \mathbb{T}_{t+1}$,
%


\begin{equation}\label{eq:app_D_10}
    \begin{split}
        \text{R.H.S. of (\ref{eq:app_D_9})} \leq& \mathbb{E}\Big[ \sum^{T}_{s=t} u_{i_{s}}(\hat{x}_{i_{s}}, \sigma_{i_{s}}(\hat{x}_{i_{s}}, \hat{\bm{x}}_{-i_{s}})) \\
        &+\sum_{s=t}^{T-1}\rho_{i_{s}}(\hat{x}_{i_{s}}, \hat{\bm{x}}_{-i_{s}}) \\
        &+ \gamma_{i_{T}} (\hat{x}_{i_{T}}, \hat{\bm{x}}_{-i_{T}}) \Big| \sigma_{i}\Big]_{\hat{x}_{i_{t}}=x_{i_{t}}}\\
        =&\mathbb{E}\Big[ \sum^{\tau}_{s=t} u_{i_{s}}(\hat{x}_{i_{s}}, \sigma_{i_{s}}(\hat{x}_{i_{s}}, \hat{\bm{x}}_{-i_{s}})) \\
        &+\sum_{s=t}^{\tau-1}\rho_{i_{s}}(\hat{x}_{i_{s}}, \hat{\bm{x}}_{-i_{s}}) \\
        &+ \rho_{i_{\tau}} + \sum_{k=\tau+1}^{T}u_{k} + \sum^{T-1}_{k=\tau+1} \rho_{k}\\
        &+ \gamma_{i_{T}} (\hat{x}_{i_{T}}, \hat{\bm{x}}_{-i_{T}}) \Big| \sigma_{i}\Big]_{\hat{x}_{i_{t}}=x_{i_{t}}}
    \end{split}
\end{equation}

From the construction of $\beta_{i_{t}}$ in (\ref{eq:beta_characterization}) and Lemma \ref{lemma:monotonicity_continuing_value}, we have,

$$
\begin{aligned}
\text{E.H.S. of \ref{eq:app_D_10} } \leq & \mathbb{E}\Big[ \sum^{\tau}_{s=t} u_{i_{s}}(\hat{x}_{i_{s}}, \sigma_{i_{s}}(\hat{x}_{i_{s}}, \hat{\bm{x}}_{-i_{s}}))\\
&+ \gamma_{i_{\tau}}(\hat{x}_{i_{t}}, \hat{\bm{x}}_{-i_{t}}) + \beta_{i_{\tau}}(\tau)\Big| \sigma_{i}\Big]_{\hat{x}_{i_{t}}= x_{i_{t}}}\\
\leq& \sup_{\tau_{i}\in \mathbb{T}_{t}}\Big\{ \mathbb{E}\Big[ \sum^{\tau}_{s=t} u_{i_{s}}(\hat{x}_{i_{s}}, \sigma_{i_{s}}(\hat{x}_{i_{s}}, \hat{\bm{x}}_{-i_{s}}))\\
&+ \gamma_{i_{\tau}}(\hat{x}_{i_{t}}, \hat{\bm{x}}_{-i_{t}}) + \beta_{i_{\tau}}(\tau)\Big| \sigma_{i}\Big]_{\hat{x}_{i_{t}}= x_{i_{t}}} \Big\},
\end{aligned}
$$
which implies 


%
\begin{equation}\label{eq:app_D_11}
    J^{\bar{S}}_{i_{t}}(x_{i_{t}}; \tilde{x}_{i_{t}}, \sigma_{i}) \leq J^{\bar{S}}_{i_{t}}(x_{i_{t}}; x_{i_{t}}, \sigma_{i}).
\end{equation}
Combining (\ref{eq:app_D_5}) and (\ref{eq:app_D_11}), we can show that (\ref{eq:def_PBNE_HI_one_shot}) is established.

\qed

\begin{lemma}\label{lemma:monotonicity_continuing_value}
Suppose Assumptions \ref{assp:single_crossing} hold. Then, $g_{i_{t}}(x_{i_{t}}; \sigma_{i})$ is non-decreasing in $x_{i_{t}}$, for all $t\in\mathbb{T}$.
\end{lemma}
\proof{}

Define 
\begin{equation}\label{eq:app_B_1}
    \begin{split}
        M_{i_{t}}(x_{i_{t}};\tilde{x}_{i_{t}}, \sigma_{i})\equiv& \mathbb{E}\Big[u_{i_{t+1}}(\hat{x}_{i_{t+1}}, \sigma_{i_{t+1}}(\hat{\bm{x}}_{t+1})) + \gamma_{i_{t+1}}(\hat{\bm{x}}_{t+1}) \\
        &+ \beta_{i_{t+1}}(t+1) \Big| \tilde{\alpha}_{i_{t}}, \sigma_{i} \Big] + \rho_{i_{t}}(\tilde{x}_{i_{t}}, \bm{x}_{-i_{t}}) \\
        &- \gamma_{i_{t}}(\tilde{x}_{i_{t}}, \bm{x}_{-i_{t}}),
    \end{split}
\end{equation}
with $M_{i_{t}}(x_{i_{t}};\sigma_{i}) = M_{i_{t}}(x_{i_{t}};x_{i_{t}}, \sigma_{i})$, such that $M_{i_{t}}(x_{i_{t}};\sigma_{i}) - \beta_{i_{t}}(t)$ is the change in agent $i_{t}$'s expected payoffs between when he continues and plans to stop at $t+1$ and he stops immediately at $t$.
Then, agent $i_{t}$'s period-$t$ interim expected payoff can be represented in terms of $M_{i_{t}}$ as follows:
\begin{equation}
    \begin{split}
        C_{i_{t}}(x_{i_{t}},\tau_{i_{t}}; \sigma_{i}) = \mathbb{E}\Big[\sum^{\tau_{i_{t}}-1}_{s=t}M_{i_{s}}(\hat{x}_{i_{s}};\sigma_{i}) -\beta_{i_{s}}(s)   \Big| \sigma_{i}\Big] \\
        &+ C_{i_{t}}(x_{i_{t}},t;\sigma_{i}).
    \end{split}
\end{equation}
Hence, $g_{i_{t}}$ in (\ref{eq:continuing_value}) can be rewritten as follows:
\begin{equation}\label{eq:app_B_3}
    \begin{aligned}
g_{i_{t}}(x_{i_{t}}; \sigma_{i}) =& \sup_{\tau_{i_{t}}\in \mathbb{T}_{t+1}} \mathbb{E}\Big[ \sum^{\tau_{i_{t}}-1}_{s=t}M_{i_{s}}(\hat{x}_{i_{s}};\sigma_{i}) -\beta_{i_{s}}(s)  \Big| \sigma_{i}\Big] \\
&+ M_{i_{t}}(x_{i_{t}};\sigma_{i}).
\end{aligned}
\end{equation}
%
Assumption \ref{assp:single_crossing} implies that $M_{i_{t}}$ is non-decreasing in $x_{i_{t}}$, for all $t\in\mathbb{T}$.
Hence, (\ref{eq:app_B_3}) leads to that $g_{i_{t}}$ is non-decreasing in $x_{i_{t}}$, for all $t\in\mathbb{T}$.


\endproof





\subsection{Proof of Lemma \ref{lemma:first_order} }\label{app:lemma_first_order}

Due to the dynamics of agent's preference, current-period reporting of preference will influence the distribution of future preferences of that agent.

Let $x_{i_{t}}\in X_{i_{t}}$ and $x_{i_{t+1}}\in X_{i_{t+1}}$ be any realized preferences in any two adjacent periods.
From (\ref{eq:dynamic_representation_1}), we have that there exists $\omega_{i_{t+1}}\in (0,1)$ such that
\begin{equation}\label{eq:app_F_1}
    x_{i_{t+1}} = \kappa_{i_{t+1}}(x_{i_{t}}, \sigma_{i_{t}}(x_{i_{t}}, \bm{x}_{-i_{t}}), \omega_{i_{t+1}}).
\end{equation}
Applying the envelope theorem in PBNE-HI mechanism yields the following:
\begin{equation}
    \frac{\partial x_{i_{t+1}}  }{\partial \ell  }\Big|_{\ell = x_{i_{t}}}= \frac{\partial \kappa_{i_{t+1}}(\ell, \sigma_{i_{t}}(x_{i_{t}}, \bm{x}_{-i_{t}}), \omega_{i_{t+1}})  }{  \partial \ell}\Big|_{\ell = x_{i_{t}}}.
\end{equation}
From the definition of $\kappa_{i_{t}}$ in (\ref{eq:dynamic_representation_0}), we have
\begin{equation}
    \begin{split}
        \frac{\partial x_{i_{t+1}}  }{\partial \ell  }\Big|_{\ell = x_{i_{t}}}=& \frac{\partial F^{-1}_{i_{t+1}}(\omega_{i_{t+1}}|\ell, \sigma_{i_{t}}(x_{i_{t}}, \bm{x}_{-i_{t}}))   }{  \partial \ell}\Big|_{\ell = x_{i_{t}}}\\
        =& \frac{ -\partial F_{i_{t+1}}(x_{t+1}|x_{t}, \sigma_{i_{t}}(x_{i_{t}}, \bm{x}_{-i_{t}})) /\partial \ell }{f_{i_{t+1}}(x_{i_{t+1}}|x_{i_{t}}, \sigma_{i_{t}}(x_{i_{t}}, \bm{x}_{-i_{t}}))    } \Big|_{\ell = x_{i_{t}}}.
    \end{split}
\end{equation}
%
%
Similarly, for any sequence of realizations of preferences, $<x_{t}, x_{t+1},\dots, x_{k},\dots x_{T}>$, we have, for any $t< k\leq T$,
\begin{equation}\label{eq:app_F_4}
    \begin{split}
        \frac{\partial x_{i_{k}}  }{\partial \ell }\Big|_{\ell=x_{i_{t}}} = &\prod_{k=t+1}^{s}\frac{-\partial F_{i_{k}}(\hat{x}_{i_{k}}| \ell, \sigma_{i_{k-1}}(\hat{\bm{x}}_{k-1}))}{f_{i_{k}}(\hat{x}_{i_{k}}|\hat{x}_{i_{k-1}}, \sigma_{i_{k-1}}(\hat{\bm{x}}_{k-1}))\partial \ell }\Big|_{\ell = x_{i_{k-1}} }.  
    \end{split}
\end{equation}

In any PBNE-HI mechanism $<\bm{\sigma}, \bm{\rho}, \bm{\gamma}, \bm{beta}>$, truthful reporting strategy is optimal for all agents.
Then, from the envelope theorem, we have

\begin{equation}\label{eq:app_F_0}
    \begin{split}
        &\frac{\partial J_{i_{t}}(\ell, \tau_{i_{t}}; \sigma_{i})   }{\partial \ell }\Big|_{\ell = x_{i_{t}}} \\
        = & \mathbb{E}\Big[\sum^{\tau_{i_{t}}}_{s=t} \frac{\partial u_{i_{s}}(\ell, \sigma_{i_{s}}(\hat{\bm{x}}_{i_{s}}))  }{\partial \ell  }\Big|_{\ell = \hat{x}_{i_{s}}}\frac{\partial  \hat{x}_{i_{s}}}{\partial \ell}\Big|_{\ell = x_{i_{t}}}\Big]\\
        =& -\mathbb{E}_{t}\Big[\sum^{\tau_{i_{t}}}_{s=t}\frac{\partial u_{i_{s}}(\ell, \sigma_{i_{s}}( \hat{\mathbf{x}}_{s})  )  }{\partial \ell}|_{\ell = \hat{x}_{i_{s}}}\\
        &\prod_{k=t+1}^{s}\frac{-\partial F_{i_{k}}(\hat{x}_{i_{k}}| \ell, \sigma_{i_{k-1}}(\hat{\bm{x}}_{k-1}))}{f_{i_{k}}(\hat{x}_{i_{k}}|\hat{x}_{i_{k-1}}, \sigma_{i_{k-1}}(\hat{\bm{x}}_{k-1}))\partial \ell }\Big|_{\ell = x_{i_{k-1}} } \Big| \sigma_{i} \Big].
    \end{split}
\end{equation}

\qed

\subsection{Proof of Theorem \ref{thm:necessary_sufficient} }\label{app:thm_necessary_sufficient}

It is easy to verify that the constructions of $\phi^{S}_{i_{t}}$ and $\phi^{\bar{S}}_{i_{t}}$ in (\ref{eq:nece_sufficent_S}) and (\ref{eq:nece_sufficent_nonS}), respectively, satisfy the sufficient conditions in Proposition \ref{prop:PBNE-HI_sufficient}.
Hence, the sufficiency can be proved in the similar way as in Appendix (\ref{app:prop_PBNE-HI_sufficient}), so we omit it here and focus on the necessity part.

\subsubsection{Optimal stopping rule calls for stopping: x$_{i_{t}} \leq \epsilon_{i}(t)$}

Let $\tilde{x}_{i_{t}}\in X_{i_{t}}$. 
Without loss of generality, suppose $\tilde{x}_{i_{t}}\leq x_{i_{t}}$.
Let $x^{1}_{i_{t}}\in X_{i_{t}}$.
To simplify the notation, we omit other agents' preferences unless otherwise stated.
Since the mechanism $<\bm{\sigma}, \bm{\rho}, \bm{\gamma}, \bm{\beta}>$ is PBNE-HI, we have
\begin{equation}\label{eq:app_G_1}
    \begin{split}
        u_{i_{t}}(x_{i_{t}}, &\sigma_{i}(x_{i_{t}}))+ \gamma_{i_{t}}(x_{i_{t}})+\beta_{i_{t}}(t)\\ &\geq  u_{i_{t}}(x_{i_{t}}, \sigma_{i}(\hat{x}_{i_{t}}))+ \gamma_{i_{t}}(\hat{x}_{i_{t}}) + \beta_{i_{t}}(t).
    \end{split}
\end{equation}

Define 
$$
\begin{aligned}
H^{S}_{i_{t}}(x^{1}_{i_{t}}) \equiv \sup_{x\in X_{i_{t}}}\Big\{ u_{i_{t}}(x^{1}_{i_{t}}, \sigma_{i}(x))+ \gamma_{i_{t}}(x)  \Big\}.
\end{aligned}
$$

Clearly, PBNE-HI implies that 
$$
x^{1}_{i_{t}}\in \arg_{x\in X_{i_{t}}}\sup \Big\{  u_{i_{t}}(x^{1}_{i_{t}}, \sigma_{i}(x))+ \gamma_{i_{t}}(x) \Big\}.
$$

Then, the differentiability implies
\begin{equation}\label{eq:app_G_2}
    \begin{aligned}
        H^{S}_{i_{t}}(x_{i_{t}}) - H^{S}_{i_{t}}(\tilde{x}_{i_{t}}) = \int^{x_{i_{t}}}_{\tilde{x}} \frac{\partial H^{S}_{i_{t}}(\ell)  }{ \partial \ell } d \ell.
    \end{aligned}
\end{equation}
From the envelope theorem, we have
\begin{equation}\label{eq:app_G_3}
    \begin{split}
        \frac{\partial H^{S}_{i_{t}}(\ell)  }{ \partial \ell }\Big|_{\ell = x_{i_{t}}}=& \frac{\partial}{ \partial \ell} \Big[ u_{i_{t}}(\ell, \sigma_{i_{t}}(x_{i_{t}}))+ \gamma_{i_{t}}(x_{i_{t}})) \Big] \Big|_{\ell = x_{i_{t}}}\\
        =& \frac{\partial}{ \partial \ell} \Big[ u_{i_{t}}(\ell, \sigma_{i_{t}}(x_{i_{t}}))\Big] \Big|_{\ell = x_{i_{t}}}\\
        =& B_{i_{s}}(x_{i_{t}};t).
    \end{split}
\end{equation}
Hence, 
$$
\begin{aligned}
\phi^{S}_{i_{t}}(x_{i_{t}}) - \phi^{S}_{i_{t}}(\tilde{x}_{i_{t}}) = &  H^{S}_{i_{t}}(x_{i_{t}}) - H^{S}_{i_{t}}(\tilde{x}_{i_{t}})\\
=& u_{i_{t}}(x_{i_{t}}, \sigma_{i_{t}}(x_{i_{t}})) + \gamma_{i_{t}}(x_{i_{t}}) \\
&- u_{i_{t}}(\tilde{x}_{i_{t}}, \sigma_{i_{t}}(\tilde{x}_{i_{t}})) - \gamma_{i_{t}}(\tilde{x}_{i_{t}})\\
\end{aligned}
$$

The definition of distance gives
$$
\begin{aligned}
d^{S}_{i_{t}}(x_{i_{t}}, \tilde{x}_{i_{t}};\sigma_{i})=& u_{i_{t}}(x_{i_{t}}, \sigma_{i_{t}}(x_{i_{t}})) - u_{i_{t}}(\tilde{x}_{i_{t}}, \sigma_{i_{t}}(x_{i_{t}}))\\
=& u_{i_{t}}(x_{i_{t}}, \sigma_{i_{t}}(x_{i_{t}})) - u_{i_{t}}(\tilde{x}_{i_{t}}, \sigma_{i_{t}}(\tilde{x}_{i_{t}}))  \\
&+ u_{i_{t}}(\tilde{x}_{i_{t}}, \sigma_{i_{t}}(\tilde{x}_{i_{t}})) - u_{i_{t}}(\tilde{x}_{i_{t}}, \sigma_{i_{t}}(x_{i_{t}}))\\
\geq& u_{i_{t}}(x_{i_{t}}, \sigma_{i_{t}}(x_{i_{t}})) - u_{i_{t}}(\tilde{x}_{i_{t}}, \sigma_{i_{t}}(\tilde{x}_{i_{t}}))\\
&- \gamma_{i_{t}}(\tilde{x}_{i_{t}}) + \gamma_{i_{t}}(x_{i_{t}})\\
=& \phi^{S}_{i_{t}}(x_{i_{t}}) - \phi^{S}_{i_{t}}(\tilde{x}_{i_{t}}).
\end{aligned}
$$

\subsubsection{Optimal stopping rule calls for continuing: $x_{i_{t}}\geq \epsilon_{i_{t}}(t)$:}

Similarly, PBNE-HI imples the following:
$$
\begin{aligned}
u_{i_{t}}(x_{i_{t}}, \sigma_{i_{t}}(&x_{i_{t}}))+ \gamma_{i_{t}}(x_{i_{t}}) + g_{i_{t}}(x_{i_{t}}; \sigma_{i})\\
\geq& u_{i_{t}}(x_{i_{t}}, \sigma_{i_{t}}(\tilde{x}_{i_{t}}))+ \gamma_{i_{t}}(\tilde{x}_{i_{t}}) + g_{i_{t}}(x_{i_{t}};\tilde{x}_{i_{t}}, \sigma_{i}).
\end{aligned}
$$
Define 
$$
\begin{aligned}
H^{\bar{S}}_{i_{t}}(x^{1}_{i_{t}}) \equiv \sup_{x\in X_{i_{t}}}\Big\{ u_{i_{t}}(x_{i_{t}}, \sigma_{i_{t}}(x))+ \gamma_{i_{t}}(x) + g_{i_{t}}(x_{i_{t}}; x, \sigma_{i})  \Big\}.
\end{aligned}
$$
PBNE-HI implies
$$
x^{1}_{i_{t}}\in \arg_{x\in X_{i_{t}}}\Big\{ u_{i_{t}}(x_{i_{t}}, \sigma_{i_{t}}(x))+ \gamma_{i_{t}}(x) + g_{i_{t}}(x_{i_{t}}; x, \sigma_{i})    \Big\}.
$$

After applying the envelope to $H^{\bar{S}}_{i_{t}}$, we obtain the following:
\begin{equation}\label{eq:app_thm_1}
\begin{split}
        &\phi^{\bar{S}}_{i_{t}}(\tilde{x}_{i_t}|\sigma_{i})- \phi^{\bar{S}}_{i_{t}}(x_{i_{t}}|\sigma_{i})= H^{\bar{S}}_{i_{t}}(\tilde{x}_{i_{t}})- H^{\bar{S}}_{i_{t}}(x_{i_{t}})\\
        =& u_{i_{t}}(x_{i_{t}}, \sigma_{i_{t}}(x_{i_{t}}))+ \gamma_{i_{t}}(x_{i_{t}}) + g_{i_{t}}(x_{i_{t}}; \sigma_{i})\\
&- u_{i_{t}}(\tilde{x}_{i_{t}}, \sigma_{i_{t}}(\tilde{x}_{i_{t}})) -\gamma_{i_{t}}(\tilde{x}_{i_{t}}) -g_{i_{t}}(\tilde{x}_{i_{t}};\sigma_{i}).
    \end{split}
\end{equation}
By expanding $g_{i_{t}}$, we have
\begin{equation}\label{eq:app_thm_2}
    \begin{split}
        &\text{R.H.S. of (\ref{eq:app_thm_1})} = \sup_{\tau_{i_{t}}\in \mathbb{T}_{t+1}}\Big\{\mathbb{E}\Big[\sum^{\tau_{i}}_{s=t} u_{i_{s}}(\hat{x}_{i_{s}}, \sigma_{i_{s}}(\hat{x}_{i_{s}}, \hat{\bm{x}}_{-i_{s}})) \\
        &+ \gamma_{i_{\tau_{i_{t}}}}(\hat{x}_{ i_{\tau_{i_{t}}}})\Big| \sigma_{i}\Big]\Big|_{\hat{x}_{i_{t}}= \tilde{x}_{t}} +\beta_{i_{\tau_{i_{t}} }} (\tau_{i_{t}})\\
        &+ \mathbb{E}\Big[\sum_{s=t}^{\tau_{i_{t}}-1}\rho_{i_{s}}(\hat{x}_{i_{s}})  \Big| \sigma_{i}\Big]\Big|_{\alpha_{i_{t}}(\tilde{x}_{i_{t}})= \tilde{x}_{t}} \Big\}\\
        &- \sup_{\tau_{i_{t}}\in \mathbb{T}_{t+1}}\Big\{\mathbb{E}\Big[\sum^{\tau_{i}}_{s=t} u_{i_{s}}(\hat{x}_{i_{s}}, \sigma_{i_{s}}(\hat{x}_{i_{s}}, \hat{\bm{x}}_{-i_{s}})) + \gamma_{i_{\tau_{i_{t}}}}(\hat{x}_{ i_{\tau_{i_{t}}}})\Big| \sigma_{i}\Big]\Big|_{\hat{x}_{i_{t}}= x_{t}} \\
        &+\beta_{i_{\tau_{i_{t}} }} (\tau_{i_{t}}) + \mathbb{E}\Big[\sum_{s=t}^{\tau_{i_{t}}-1}\rho_{i_{s}}(\hat{x}_{i_{s}})  \Big| \sigma_{i}\Big]\Big|_{\alpha_{i_{t}}(x_{i_{t}})= x_{t}} \Big\}.
    \end{split}
\end{equation}
Then we have the following:
\begin{equation}\label{eq:app_thm_3}
    \begin{split}
        &\text{R.H.S. of (\ref{eq:app_thm_2})}\leq \sup_{\tau_{i_{t}}\in \mathbb{T}_{t+1}}\Big\{\mathbb{E}\Big[\sum^{\tau_{i}}_{s=t} u_{i_{s}}(\hat{x}_{i_{s}}, \sigma_{i_{s}}(\hat{x}_{i_{s}}, \hat{\bm{x}}_{-i_{s}})) \\
        &+ \gamma_{i_{\tau_{i_{t}}}}(\hat{x}_{ i_{\tau_{i_{t}}}})\Big| \sigma_{i}\Big]\Big|_{\hat{x}_{i_{t}}= \tilde{x}_{t}}  \\
        &+ \mathbb{E}\Big[\sum_{s=t}^{\tau_{i_{t}}-1}\rho_{i_{s}}(\hat{x}_{i_{s}})  \Big| \sigma_{i}\Big]\Big|_{\alpha_{i_{t}}(\tilde{x}_{i_{t}})= \tilde{x}_{t}} \\
        &- \mathbb{E}\Big[\sum^{\tau_{i}}_{s=t} u_{i_{s}}(\hat{x}_{i_{s}}, \sigma_{i_{s}}(\hat{x}_{i_{s}}, \hat{\bm{x}}_{-i_{s}})) + \gamma_{i_{\tau_{i_{t}}}}(\hat{x}_{ i_{\tau_{i_{t}}}})\Big| \sigma_{i}\Big]\Big|_{\hat{x}_{i_{t}}= x_{t}}\\
        & - \mathbb{E}\Big[\sum_{s=t}^{\tau_{i_{t}}-1}\rho_{i_{s}}(\hat{x}_{i_{s}})  \Big| \sigma_{i}\Big]\Big|_{\alpha_{i_{t}}(x_{i_{t}})= x_{t}} \Big\}.
    \end{split}
\end{equation}

PBNE-HI implies

\begin{equation}\label{eq:app_thm_4}
    \begin{split}
        &\text{R.H.S. of (\ref{eq:app_thm_3})}\leq \sup_{\tau_{i_{t}}\in \mathbb{T}_{t+1}}\Big\{\mathbb{E}\Big[\sum^{\tau_{i}}_{s=t} u_{i_{s}}(\hat{x}_{i_{s}}, \sigma_{i_{s}}(\hat{x}_{i_{s}}, \hat{\bm{x}}_{-i_{s}})) \\
        &+ \gamma_{i_{\tau_{i_{t}}}}(\hat{x}_{ i_{\tau_{i_{t}}}})\Big| \sigma_{i}\Big]\Big|_{\hat{x}_{i_{t}}= \tilde{x}_{t}} \\
        &+ \mathbb{E}\Big[\sum_{s=t}^{\tau_{i_{t}}-1}\rho_{i_{s}}(\hat{x}_{i_{s}})  \Big| \sigma_{i}\Big]\Big|_{\alpha_{i_{t}}(\tilde{x}_{i_{t}})= \tilde{x}_{t}} \\
        &- \mathbb{E}\Big[\sum^{\tau_{i}}_{s=t} u_{i_{s}}(\hat{x}_{i_{s}}, \sigma_{i_{s}}(\hat{x}_{i_{s}}, \hat{\bm{x}}_{-i_{s}})) + \gamma_{i_{\tau_{i_{t}}}}(\hat{x}_{ i_{\tau_{i_{t}}}})\Big|\tilde{x}_{i_{t}}, \sigma_{i}\Big]\Big|_{\hat{x}_{i_{t}}= x_{t}}\\
        &- \mathbb{E}\Big[\sum_{s=t}^{\tau_{i_{t}}-1}\rho_{i_{s}}(\hat{x}_{i_{s}})  \Big| \sigma_{i}\Big]\Big|_{\alpha_{i_{t}}(x_{i_{t}})= \tilde{x}_{t}} \Big\}\\
        \leq&  \sup_{\tau_{i_{t}}\in \mathbb{T}_{t+1}}\Big\{ \mathbb{E}\Big[\sum^{\tau_{i}}_{s=t} u_{i_{s}}(\hat{x}_{i_{s}}, \sigma_{i_{s}}(\hat{x}_{i_{s}}, \hat{\bm{x}}_{-i_{s}})) \\
        &+ \gamma_{i_{\tau_{i_{t}}}}(\hat{x}_{ i_{\tau_{i_{t}}}})\Big| \sigma_{i}\Big]\Big|_{\hat{x}_{i_{t}}= \tilde{x}_{t}}\\
        &- \mathbb{E}\Big[\sum^{\tau_{i}}_{s=t} u_{i_{s}}(\hat{x}_{i_{s}}, \sigma_{i_{s}}(\hat{x}_{i_{s}}, \hat{\bm{x}}_{-i_{s}})) + \gamma_{i_{\tau_{i_{t}}}}(\hat{x}_{ i_{\tau_{i_{t}}}})\Big|\tilde{x}_{i_{t}}, \sigma_{i}\Big]\Big|_{\hat{x}_{i_{t}}= x_{t}}\Big\}\\
        &+\max_{\tau\in \mathbb{T}_{\tau_{i_{t}}}}\beta_{\tau}(\tau).
    \end{split}
\end{equation}

Then, from the definition of distance function $d^{\bar{S}}_{i_{t}}$, we have
$$
\begin{aligned}
&\text{R.H.S of (\ref{eq:app_thm_4})} =  \sup_{\tau_{i_{t}}\in \mathbb{T}_{t+1}}\Big\{d^{\bar{S}}_{i_{t}}(\tilde{x}_{i_{t}}, x_{i_{t}}, \tau_{i_{t}}; \sigma_{i})\Big\}.
\end{aligned}
$$
Hence, we obtain $\sup_{\tau_{i_{t}}\in \mathbb{T}_{t+1}}\Big\{ d^{S}_{i_{t}}(\tilde{x}_{i_{t}}, x_{i_{t}}; \tau_{i_{t}}, \sigma_{i} ) \Big\} \geq \phi^{\bar{S}}_{i_{t}}(\tilde{x}_{i_t}|\sigma_{i})- \phi^{\bar{S}}_{i_{t}}(x_{i_{t}}|\sigma_{i})$.

\qed

\subsection{Proof of Proposition \ref{prop:design_beta_principle} }\label{app:prop_design_beta_principle}

\subsubsection{Statement (1) and (2)}
Fix a $\sigma_{i}$.
If the mechanism $<\bm{\sigma}, \bm{\rho}, \bm{\gamma}, \bm{\beta}>$ is PBNE-HI, then any posted price sequence $r_{i}$ produced by $\beta_{i}$ with $\epsilon_{i}$ satisfies the construction of $\beta_{i}$ given in (\ref{eq:beta_characterization}) due to Theorem \ref{thm:necessary_sufficient}.

For the converse direction, let $\beta'_{i}$ be the rule for agent $i$ in a PBNE-HI mechanism, with $r'_{i}=\beta'_{i}(<0,\dots,T>)$ and the corresponding unique threshold rule $\epsilon_{i}$.
Suppose $r'_{i}\not\in R_{i}[\sigma_{i}, \epsilon_{i}]$, e.g., 

$$
\begin{aligned}
r_{i_{t}}\equiv& \Big\{r_{i}\in\mathbb{R}^{T}:\mathbb{E}\Big[ r_{i_{t}} + \sum_{s=t}^{T-1}\big(\phi^{S,\bar{S}}_{i_{s}}(\hat{x}_{i_{s}}\vee \epsilon_{i}(s)) \\
&- \phi^{S,\bar{S}}_{i_{s+1}}(\hat{x}_{i_{s+1}}\vee \epsilon_{i}(s+1)) \big) \Big| x_{i_{t}} =  \epsilon_{i}(t),\sigma_{i}\Big] =0  \Big\},
\end{aligned}
$$
%
%
which implies that there exists some $q>0$, such that
\begin{equation}\label{eq:app_prop_design_1}
    \beta_{i_{t}}(t) + q= g_{i_{t}}(x_{i_{t}};\sigma_{i}).
\end{equation}

Suppose agent $i$'s period-$t$ realized preference is $x_{i_{t}}\leq \epsilon_{i}(t)$. Then truthful reporting and stopping immediately gives agent $i$ the expected payoff 
$$
\begin{aligned}
C_{i_{t}}(x_{i_{t}}, t; \sigma_{i})=& \sum^{t-1}_{s=0}u_{i_{s}}(x_{i_{s}}, \sigma_{i_{s}}(x_{i_{s}}, \bm{x}_{-i_{s}})) + \rho_{i_{s}}( x_{i_{s}}, \bm{x}_{-i_{s}}) ) \\
& + u_{i_{t}}(x_{i_{t}},  \sigma_{i_{t}}(x_{i_{t}}, \bm{x}_{-i_{t}}) ) + \gamma_{i_{t}} (x_{i_{t}}, \bm{x}_{-i_{t}})) +\beta_{i_{t}}(t).
\end{aligned}
$$
From (\ref{eq:app_prop_design_1}), we have
$$
\begin{aligned}
C_{i_{t}}(x_{i_{t}}, t; \sigma_{i})\leq& \sum^{t-1}_{s=0}u_{i_{s}}(x_{i_{s}}, \sigma_{i_{s}}(x_{i_{s}}, \bm{x}_{-i_{s}})) + \rho_{i_{s}}( x_{i_{s}}, \bm{x}_{-i_{s}}) ) \\
& + u_{i_{t}}(x_{i_{t}},  \sigma_{i_{t}}(x_{i_{t}}, \bm{x}_{-i_{t}}) ) + \gamma_{i_{t}} (x_{i_{t}}, \bm{x}_{-i_{t}})) + g_{i_{t}}(x_{i_{t}};\sigma_{i}),
\end{aligned}
$$
which incentivizes agent $i_{t}$ to continue, which contradits the optimality of stopping rule.

Statement (2) can be proved by setting $\beta_{i_{t}}(t)=0$ for all $i_{t}\in I_{t}$, $t\in \mathbb{T}$.

\subsubsection{Statement (3)}

To make all agents non-stop at all intermediate periods $t\in \mathbb{T}_{0,T-1}$, the posted-price payment rule $\bm{\beta}$ has to impose the threshold function $\epsilon_{i}(t) = \bar{x}_{i_{t}}$, for all $i_{t}\in I_{t}$, $t\in\mathbb{T}$, i.e., $\{g_{i_{t}}(\bar{x}_{i_{t}};\sigma_{i})\}_{t\in \mathbb{T}}\in R_{i}[\sigma_{i}, \bar{x}_{i_{t}}]$, for all $i_{t}\in I_{t}$, $t\in\mathbb{T}$. We can prove this in the similar way as that for statements (1) and (2).

Suppose the CP uses a global rule $\beta_{i}=\beta$, for all $i\in I$, with $\epsilon_{i}(t) = \bar{x}_{i_{t}}$, for all $i_{t}\in I_{t}$, $t\in\mathbb{T}$, which produces a sequence of posted price $r=<r_{0}, \dots, r_{T}>$.
Then, the mechanism is PBNE-HI if and only if $\{g_{i_{t}}(\bar{x}_{i_{t}};\sigma_{i})\}_{t\in \mathbb{T}}\in R_{i}[\sigma_{i}, \bar{x}_{i_{t}}]$, or equivalently,

$$
\begin{aligned}
r_{t}\equiv \Big\{r_{i}\in\mathbb{R}^{T}:&\mathbb{E}\Big[ r_{i_{t}} + \sum_{s=t}^{T-1}\big(\phi^{S,\bar{S}}_{i_{s}}(\bar{x}_{i_{s}}) \\
&- \phi^{S,\bar{S}}_{i_{s+1}}(\bar{x}_{i_{s+1}}) \big) \Big| x_{i_{t}} =  \bar{x}_{i_{t}},\sigma_{i}\Big] =0  \Big\}.
\end{aligned}
$$




\qed

\bibliographystyle{IEEEtran}
\bibliography{root}

\end{document}